\documentclass[12pt,preprint]{aastex}

\newcommand{\che} {\log\ ({\rm C/He})}

\newcommand{\msun} {$M_{\odot}$}

\newcommand{\halpha} {H$\alpha$}

\newcommand{\Te} {T_{\rm eff}}
\newcommand{\logg} {\log g}
\newcommand{\nh} {\log\ ({\rm H/He})}

\begin{document}

\title{Detailed Spectroscopic and Photometric Analysis of DQ White Dwarfs}

\author{P. Dufour, P. Bergeron, and G. Fontaine}
\affil{D\'{e}partement de Physique, Universit\'{e} de Montr\'{e}al,
C.P. 6128, Succ. Centre-Ville, Montr\'{e}al, Qu\'{e}bec, Canada H3C 3J7} 
\email{dufourpa@astro.umontreal.ca, bergeron@astro.umontreal.ca, 
fontaine@astro.umontreal.ca}

\begin{abstract}

We present an analysis of spectroscopic and photometric observations
of cool DQ white dwarfs based on improved model atmosphere
calculations. In particular, we revise the atmospheric parameters of
the trigonometric parallax sample of Bergeron, Leggett, \& Ruiz, and
discuss the astrophysical implications on the temperature scale and
mean mass, as well as the chemical evolution of these stars. We also
analyze 40 new DQ stars discovered in the first data release of the
Sloan Digital Sky Survey. Our analysis confirms that effective
temperatures derived from model atmospheres including carbon are
significantly lower than the temperatures obtained from pure helium
models. Similarly the mean mass of the trigonometric parallax sample,
$\langle M \rangle=0.62$ \msun, is significantly lower than that
obtained from pure helium models, $\langle M \rangle$= 0.73 \msun, and
more consistent with the spectroscopic mean mass of DB stars, $\langle
M \rangle$= 0.59 \msun, the most likely progenitors of DQ white
dwarfs. We find that DQ stars form a remarkably well defined sequence
in a carbon abundance versus effective temperature diagram; below
$\Te\sim$10,000 K, carbon pollution decreases monotonically with
decreasing effective temperature. Improved evolutionary models
including diffusion and connecting to the PG~1159 phase are used to
infer a typical value for the thickness of the helium layer $M_{\rm
He}/M_{\star}$ between $10^{-3}$ and $10^{-2}$, compatible with the
predictions of post-AGB models. Several DQ stars in our sample,
however, show larger than average carbon abundances. We argue that
these DQ stars are all massive white dwarfs, and could represent the
high mass tail of the white dwarf mass distribution, with their hotter
counterparts corresponding to the hot DQ stars reported recently by
Liebert et al. The number distribution of DQ white dwarfs as a
function of effective temperature clearly shows a sudden drop at about
$\Te\sim$7000 K and an abrupt cutoff at $\Te\sim$6000 K. The existence
of this cutoff is now statistically more significant with the addition
of the SDSS stars. The physical mechanism responsible for this cutoff
is still unknown even though it is believed to be somehow related to
the existence of the so-called C$_2$H stars at lower temperatures.

\end{abstract}


\section{INTRODUCTION}

White dwarfs showing spectroscopic traces of carbon, either as neutral
carbon lines or molecular C$_2$ Swan bands, are collectively known as
DQ stars. Past analyses have shown that these stars have helium-rich
atmospheres with carbon abundances ranging from $\che=-7$ to $-2$, as
determined from optical or ultraviolet (mainly $IUE$) spectroscopic
observations \citep{bues73,grenfell74,koester82,wegner84,weidemann95}.
The presence of carbon in the atmosphere of these objects has been
successfully explained by a model in which carbon diffusing upward
from the core is brought to the photosphere by the deep helium
convection zone \citep{pelletier86}. This model was shown to reproduce
the observed carbon abundance distribution as a function of effective
temperature fairly well for a helium layer thickness of $\log q({\rm
He})\equiv \log M_{\rm He}/M_{\star}\sim -3.75$, a value at odds with those
obtained from evolutionary models of post-AGB stars that predict much
thicker helium layers of the order of $\log q({\rm He})\sim -2$
\citep{dantona79,iben84,iben85}.

More recently, \citet{brl97} and \citet[][hereafter referred to as BRL
and BLR, respectively]{blr01} presented a photometric and
spectroscopic analysis of a large sample of cool white dwarfs aimed at
improving our understanding of the chemical evolution of their outer
layers. Several DQ stars were included in both studies and analyzed
under the assumption of pure helium compositions, the only models
available at that time.  The location in a mass versus effective
temperature diagram for the 17 DQ stars\footnote{Two additional DQ
stars (LDS 275A and ESO 267-110) have been included here with respect
to the BLR analysis by making use of new trigonometric parallax measurements
reported in \S~\ref{observation}.} with measured trigonometric
parallaxes is reproduced here in Figure
\ref{fg:f1} together with that of other hydrogen- and
helium-rich atmosphere white dwarfs taken from BLR. One of the
outstanding problems raised by BRL and BLR was the complete absence of
cool DQ stars below $\Te\lesssim6500$~K. With an estimated age of the
local Galactic disk of roughly 8 Gyr according to
\citet{lrb98}, DQ stars should exist at much lower
effective temperatures according to the results shown in Figure
\ref{fg:f1}. Even with the carbon diffusion tail receding from the
photospheric regions with decreasing effective temperature
\citep[see Fig.~1 of][]{pelletier86}, the C$_2$ Swan bands should be 
easily detectable in much cooler objects, even for the small carbon
abundances expected from evolutionary calculations (see
\S~\ref{theoretical} below). On the basis of the results shown in Figure
\ref{fg:f1}, BRL and BLR suggested that the absence of DQ stars 
was not a selection effect, but was instead related to some yet
unexplained physical mechanism.

Strong molecular absorption bands that appear to originate from some
carbon compound have also been reported in several cool white dwarfs
\citep[see, e.g.][]{schmidt95,harris03}. The detailed model atmosphere 
and photometric analysis by BRL of some of these objects reveals that
in addition to the existence of these molecular features, the infrared
fluxes are strongly reduced as a result of collision induced
absorption by molecular hydrogen due to collisions with helium
\citep[see the case of LHS 1126 analyzed by][]{bergeron94}. 
Typical hydrogen abundances of $\nh\sim -1$ were able to account for
the observed infrared flux deficiency. The simultaneous presence of
hydrogen, helium, and carbon in these white dwarfs led
\citet{schmidt95} to suggest, on the basis of a detailed chemical
equilibrium analysis of H/He/C mixtures under the physical conditions
encountered in the atmospheres of these stars, that C$_2$H is the
molecule preferentially formed in the photospheric regions. Hence the
strong unidentified molecular features could be due to the C$_2$H
molecule, and these white dwarfs have been termed C$_2$H stars,
although the lack of detailed theoretical absorption profile
calculations prevents us from confirming this interpretation.

All C$_2$H stars that have been analyzed by BRL are found at low
effective temperatures, and below the observed cutoff in the DQ
temperature distribution. The most obvious explanation proposed by BRL
to account for the sudden drop in the number of DQ stars below $\Te\sim
6500$~K is that DQ stars turn into C$_2$H stars as they cool off.
This scenario has recently been challenged by
\citet{carollo03} who reported the discovery of a DQ
star with very strong C$_2$ Swan bands \citep[see also][]{carollo02}
at an estimated effective temperature of $\sim5120$~K, well below the
empirical cutoff alluded to by BRL. Taken at face value, this result
would suggest that the cutoff is simply a selection effect and that
the DQ phenomenon persists at low effective temperatures.

Progress in our understanding of the evolution of DQ stars and C$_2$H
stars, and whether they are related to each other, must go through a
homogeneous determination of the atmospheric parameters of a large
ensemble of DQ white dwarfs. Clearly, the parameters obtained by BRL
and BLR are only approximate since these were determined from model
atmospheres with pure helium compositions. Indeed, \citet{provencal02} demonstrated the importance of
including carbon in the model atmosphere calculations for the detailed
spectroscopic and photometric analysis of the DQ star Procyon B.
Their effective temperature determination of $\Te=7740\pm 50$~K,
derived from models including carbon, was significantly cooler than
the earlier value of $\Te=8688\pm 200$~K obtained by
\citet{provencal97} on the basis of pure helium models. This large
difference was attributed to an increase of the He$^-$ free-free
opacity resulting from the additional free electrons provided by
carbon (see also \S~\ref{theoretical} below).

The natural step following the comprehensive studies of BRL and BLR is
thus to include explicitly the presence of carbon in the model
atmosphere calculations, and to reanalyze the available photometric
and spectroscopic data for DQ stars. In addition, we present a similar
analysis of the DQ stars reported in the first data
release of the Sloan Digital Sky Survey (SDSS). In
\S~\ref{theoretical}, we describe our theoretical framework including
our model atmosphere and synthetic spectrum calculations.  The
observations described in \S~\ref{observation} are then analyzed in
detail in \S~\ref{analysis}. The results are then interpreted in
\S~\ref{results} and our conclusions follow in \S~\ref{conclusion}.

\section{THEORETICAL FRAMEWORK}\label{theoretical}

The LTE model atmosphere code used in our analysis is a modified
version of that described at length in \citet{bergeron95}, which is
appropriate for pure hydrogen and pure helium atmospheric
compositions, as well as mixed hydrogen and helium compositions.
Energy transport by convection is treated within the mixing-length
theory. In order to improve upon the BRL and BLR analyses of the DQ
and eventually the DZ spectral types, we need to include metals and
molecules in our equation of state and opacity calculations. We
describe these modifications in turn.

\subsection{Equation of State}

The equation of state used in the original code included the following
species: \ion{H}{1}, \ion{H}{2}, H$^-$, H$_2$, H$_2^+$, H$_3^+$,
\ion{He}{1}, \ion{He}{2}, He$^-$, and He$_2^+$.
We now include C, N, O, Na, Mg, Si, Ca, Fe and their ions, as well as
over 20 diatomic and polyatomic molecules that can be formed from
these elements. Of particular interest for our analysis of DQ stars,
we have explicitly included C$_2$, C$_3$, CH, C$_2$H, CH$_2$,
C$_2$H$_2$, CH$_3$ and CH$_4$. We determine the partial pressure of
every species by solving the equations of chemical and ionization
equilibria with a Newton-Raphson iterative scheme similar to that
described in \citet{allard90}. The atomic partition functions are
calculated by summing explicitly over the bound states taken from
TOPBASE, while those of molecular species are obtained from the
polynomial functions tabulated by \citet{Irwin81}.

\subsection{Opacity Sources}

Cool, helium-rich atmosphere white dwarfs are characterized by high
atmospheric pressures resulting from the low opacity of helium.
Pressure broadening is therefore the dominant mechanism for line
broadening, which is treated under the impact approximation with van
der Waals broadening by neutral helium. The effect of Doppler
broadening is also taken into account with the use of a Voigt
profile. Over 700 of the strongest lines of \ion{C}{1} and \ion{C}{2}
are included explicitly in the model and synthetic spectrum
calculations. Central wavelengths of the transitions, $gf$ values,
energy levels, and damping constants (radiative, Stark, and van der
Waals, the first two being negligible under the physical conditions
encountered in cool white dwarf atmospheres) are extracted from the GFALL
linelist of R.~L.~Kurucz \footnote{See
http://kurucz.harvard.edu/LINELISTS.html}.

For the cool DQ stars ($\Te\lesssim8000$~K), the mean distance
between atoms is not much larger than the size of the atoms and
molecules. As discussed by \citet{koester82}, the resulting strong
interaction leads to the break down of the impact approximation. More
elaborate broadening theories for fitting the lines in the ultraviolet
yield better results --- in particular for the asymmetric
\ion{C}{1} $\lambda$1930 line --- even though the results are still
not entirely satisfactory. Since there are no atomic lines in the
optical spectra of cool DQ stars, the use of the impact approximation
appears justified for our study.

Carbon molecular bands are calculated with the ``Just Overlapping Line
Approximation'' \citep[JOLA;][]{golden67} following the version
described in \citet{zeidler82}. About 200 bands of the Swan and
Deslandres-d'Azambuja systems are included in both the model and
synthetic spectrum calculations. The molecular data are taken from
\citet{huber79} and the vibrational transition probability data
from \citet{cooper75}.

\subsection{Model Atmospheres and Synthetic Spectra}

Since we are only concerned in this analysis with DQ stars, only
helium and carbon are taken into account, although models including
traces of hydrogen have also been calculated. Our model grid covers a
range from $\Te=5000$ to 12,000~K in steps of 500~K, $\logg=7.5$ to
9.5 in steps of 0.5 dex, and $\che=-9.0$ to $-2.0$ in steps of 0.5
dex. Additional models with $\nh=-5.0$ and $-3.0$ have also been
calculated for the same range of effective temperatures and carbon
abundances as above but with a surface gravity fixed at $\logg=8$.
Detailed synthetic fluxes are obtained from these atmospheric
structures.

Illustrative synthetic spectra from our model grid at $\logg=8$ are
displayed in Figure \ref{fg:f2} for various values of
effective temperatures and carbon abundances. We note that comparable
line strengths can be produced by compensating a decrease in effective
temperature by a decrease in the carbon abundance. Additional
information must thus be sought to constrain the effective
temperature, such as spectrophotometric or broadband photometric
observations. We find that surface gravity has little effect on the
strength of the Swan bands. This might seem surprising at first since
the surface gravity determines the atmospheric pressure and thus the
number density of atoms, which in turn affects the population of
molecular carbon since $N(\rm C_2)$ $\propto N$(\ion{C}{1}$)^2$. But the
He$^-$ free-free opacity is proportional to the product of the number
of free electrons and neutral helium, which both increase with surface
gravity as well. The net effect is that the increase of the C$_2$ Swan
band opacity is partially masked by an increase in the He$^-$ free-free
opacity.

As discussed in the Introduction, the presence of carbon not only
produces the strong atomic and molecular absorption features shown in
Figure \ref{fg:f2}, but it also contributes to increase the
number of free electrons, and thus the He$^-$ free-free opacity. The
resulting effect of carbon on the atmospheric structure is illustrated
in Figure \ref{fg:f3} where we contrast the temperature
structure of models at $\Te=8000$~K and $\logg=8$ with (1) a pure
helium composition, (2) a carbon abundance of $\che=-5.0$ included
only in the equation of state (i.e.,~the carbon opacity is neglected),
and (3) a fully blanketed model. We can see that the main effect on
the temperature structure comes from the inclusion of carbon in the
equation of state calculation, and that the inclusion of the carbon
absorption features has only a small effect on the atmospheric
structure, particularly in the continuum forming region near
$\tau_{\rm R}\sim1$.

We have validated our model atmosphere calculations by comparing our
results with those of \citet{provencal02} and \citet{carollo03},
which both rely on the DQ models of D.~Koester. 
An analysis of the DQ star Procyon B similar to that described in 
\citet{provencal02} yields $\Te=7710$~K and $\che=-5.4$, in excellent
agreement with the results obtained by Provencal et al., namely, $\Te=7740$~K
and $\che=-5.5$. Similarly, we show in Figure \ref{fg:f4} the
optical spectrum of the DQ star GSC2U J131147.2+292348, kindly made
available to us by D. Carollo, together with our synthetic spectrum
calculated at the atmospheric parameter solution obtained by
\citet{carollo03}, $\Te=5200$~K and $\che=-5.53$ at $\logg=8$.  A
comparison of these results with those shown in Figure 2 of
\citet{carollo03} indicates that our models are in excellent 
agreement with the predictions of Koester's models.

\section{OBSERVATIONS}\label{observation}

The first sample used for this study is drawn from the BRL and BLR
analyses, which include 21 DQ stars. From this we exclude those
objects that have carbon features only in the ultraviolet
(WD~0038$+$555, WD~0806$-$661), missing or bad optical spectra
(WD~1708$-$147, WD~2254$+$076), and uncertain photometric measurements
(WD~2154$-$512). We are left with a sample of 16 DQ stars with optical
$BVRI$ and infrared $JHK$ photometry, as well as high signal-to-noise
spectroscopy showing the carbon molecular bands, all of which have
trigonometric parallax measurements. The latter are given in Table 1
of BLR, with the exception of LDS 275A (WD~0935-371; $\pi=43.7\pm1.9$
mas) and ESO 267-110 (WD~1157-462; $\pi 18.5\pm 1.9$ mas) whose
measurements come from M. T. Ruiz (2005, private
communication). Optical spectra for most of these stars have already
been displayed in Figure 20 of BRL and Figure 6 of BLR, while the
photometric data and parallax measurements can be found in Table 1 of
both papers. Data acquisition and reduction are described in these
papers as well. New optical spectroscopic observations for
WD~2311$-$068 and WD~2140+207 have been secured at the mont M\'egantic
Observatory 1.6 m telescope in 2004 September with a setup similar to
that used for the other objects.

Our second sample consists of white dwarf stars spectroscopically
identified as DQ (human ID) in the first data release of the
SDSS. Details concerning these observations are described in
\citet{kleinman04} and references therein. Photometric
measurements on the $ugriz$ system are taken from the electronic
tables of the SDSS web site
\footnote{http://www.sdss.org/dr1/products/value\_added/wdcat/dr1/wdALL.table.DR1}. 
From a total of 50 DQ stars, we exclude 8 objects showing atomic
carbon lines similar to those reported by \citet{liebert03}, and that
are hotter than the upper boundary of our model grid at
$\Te=12,000$~K.  We also exclude one peculiar white dwarf (C$_2$H
star) and another object whose spectrum is too noisy. Our SDSS sample
thus consists of 40 DQ stars. Representative spectra for these objects
are displayed in Figures 7 and 8 of \citet{harris03}.
 
\section{DETAILED ANALYSIS}\label{analysis}

\subsection{Fitting Technique}

The method used to fit the photometric data is similar to that
described at length in BRL, which we summarize here for
completeness. We first transform the magnitudes at each bandpass into
observed average fluxes $f_{\lambda}^m$ using the following equation

$$m= -2.5\log f_{\lambda}^m + c_m\ , \eqno (1)$$

\noindent where the values of the constants $c_m$ are given in BRL. The
resulting energy distributions are then fitted with the model
Eddington fluxes $H_{\lambda}^m$ --- which here depend on $\Te$,
$\logg$, and the carbon abundance, properly averaged over the same
filter bandpasses.  The average observed and model fluxes are related
by the equation

$$f_{\lambda}^m= 4\pi~(R/D)^2~H_{\lambda}^m\ ,\eqno (2)$$

\medskip
\noindent where $R/D$ is the ratio of the radius of the star to its distance 
from Earth. Our fitting procedure relies on the nonlinear
least-squares method of Levenberg-Marquardt \citep{pressetal92}, which
is based on a steepest descent method.  The value of $\chi ^2$ is
taken as the sum over all bandpasses of the difference between both
sides of equation (2), properly weighted by the corresponding
observational uncertainties. We consider only $\Te$ and the solid
angle $\pi(R/D)^2$ free parameters, and keep the carbon
abundance fixed when fitting the energy distribution. The
uncertainties of $\Te$ and the solid angle are obtained directly from
the covariance matrix of the fit.

We begin by assuming a pure helium composition and $\logg=8$ to get
initial estimates of $\Te$ and the solid angle.  The latter is
combined with the distance $D$ obtained from the trigonometric
parallax to determine the stellar radius $R$, which is then converted
into $\logg$ (or mass) using evolutionary models similar to those
described in \citet{fon01} but with C/O cores, $q({\rm He})=10^{-2}$,
and $q({\rm H})=10^{-10}$, which are representative of
helium-atmosphere white dwarfs. The whole procedure is repeated until
a convergence in $\logg$ is reached.  For stars with no trigonometric
parallax measurements available, we simply assume $\logg=8$
throughout.

Next we turn to the spectroscopic observations and determine the
carbon abundance by fitting the Swan bands at the values of $\Te$ and
$\logg$ obtained from the fit to the energy distribution. This is done
by minimizing the value of $\chi ^2$ taken as the sum over all
frequencies of the difference between the observed and model fluxes,
all frequency points being given an equal weight.  The model fluxes
are multiplied by a polynomial of second degree in $\lambda$ in order
to account for the solid angle in equation (2), as well as residual
uncertainties in the flux calibration of the optical spectra (see
below). The carbon abundance and the coefficients of this polynomial
are considered free parameters when fitting the optical spectrum, while
$\Te$ and $\logg$ are kept constant.  

New estimates of $\Te$ and $\logg$ are then obtained by fitting again
the photometric observations, but this time with models interpolated
at the carbon abundance obtained from the spectroscopic fit. The
complete procedure is repeated --- typically five iterations --- until
$\Te$, $\log g$, and the carbon abundance converge to a consistent
photometric and spectroscopic solution.

\subsection{Reappraisal of the BRL and BLR Analyses}\label{BLR}

We first begin by examining the implications of our improved model
atmosphere calculations on the atmospheric parameter determinations of
the DQ stars in the BRL/BLR sample. Illustrative fits to the
energy distribution of the DQ star LHS 43 (WD~1142$-$645) are shown in
Figure \ref{fg:f5} under the assumption of a pure helium
atmospheric composition (top panel) and with a carbon abundance
constrained from the optical spectrum (bottom panel). The quality of
the fit to the energy distribution is not improved by the inclusion of
carbon, but both the effective temperature and surface gravity have
been significantly reduced.

The importance of including carbon for determining the atmospheric
parameters of DQ stars can be appreciated by studying the results of
Figure \ref{fg:f6} where we show the effective temperature
solution obtained from fits to the energy distribution of LHS 43 for
various {\it assumed} carbon abundances. We see that as the carbon
abundance is gradually increased (starting from a pure helium
composition), the $\Te$ solution drops by several hundred degrees
until it reaches a minimum plateau. For larger carbon abundances,
the Swan bands become so strong that the $B$ and $V$ fluxes 
are significantly reduced, requiring the $\Te$ solution to rise
again in order to match the observed fluxes in these bandpasses. Of
course, such high carbon abundances yield unacceptable photometric
fits and can also be ruled out from our spectroscopic
observations. Also shown in Figure \ref{fg:f6} are the
corresponding $\chi^2$ values obtained from the photometric fits. We
see that the $\chi^2$ minimum lies very close to the solution obtained
from our iterative photometric and spectroscopic scheme, shown here as
the solid dot. This internal consistency is even better for DQ stars
with strong bands.

The final results for the 16 DQ stars taken from the BRL and BLR
analyses are reported in Table 1, where the effective temperature,
surface gravity, stellar mass, and carbon abundance are given for each
object. Quantities in parentheses represent the formal $1\sigma$
uncertainties of each fitted parameter obtained from our fitting
procedure. The corresponding fits to the optical spectra are displayed
in Figure \ref{fg:f7}. Some of the spectra shown here are
obviously not flux calibrated properly. For instance, the observed
flux in the blue portion of the spectra of WD~0856+331 and WD~0115+159
drops too steeply with respect to other stars with comparable
effective temperatures. This effect is taken into account in our
fitting procedure by allowing a polynomial correction to the model
fluxes, as discussed above.  The coefficients for this polynomial are
generally small for the remaining stars, however.

The model spectra reproduce the main C$_2$ features fairly well, most
notably the $\Delta\nu=+1$ and $0$ bands near 4740 \AA\ and 5160 \AA,
respectively, as well as the weaker $\Delta\nu=-1$ bands near 5660
\AA, when present. We note, however, that the $\Delta\nu=+1$ bands are
predicted too strong in some DQ stars (see, e.g., WD~2059$+$316 and
WD~2140+207). The origin of this discrepancy eludes us at this point.
Also, the strong feature near $\sim 4300$ \AA\ in the spectrum of
G99-37 (WD~0548$-$001) is the CH ``G'' band, which is not included
explicitly in our calculations. 

Neutral carbon lines can be observed in the spectrum of G47-18
(WD~0856+331). While our fit to the Swan bands for this star is
satisfactory, the depths of the atomic carbon lines are predicted too
shallow. The addition of hydrogen or other sources of free electrons
(metals) does not help resolve this discrepancy.  We find that a much
better fit to the carbon lines and the Swan bands can be achieved
simultaneously by increasing artificially both the effective
temperature and the carbon abundance, at the expense of being
inconsistent with our photometric solution, however. This suggests
that the balance between atomic and molecular carbon may not be
calculated properly, perhaps because of uncertainties in partition
functions or binding energies.

We compare in Figure \ref{fg:f8} the effective temperatures and
masses derived from pure helium models (i.e.,~BLR) with those obtained
here with models including carbon.  We find that our revised effective
temperatures are significantly lowered with respect to the pure helium
solutions, in agreement with the conclusions of \citet{provencal02}
for Procyon B. The inclusion of carbon in the equation of state
increases the number of free electrons and thus the contribution of
the He$^-$ free-free opacity, which in turn affects the atmospheric
structure, in particular in the continuum forming region (see
Fig.~\ref{fg:f3}). Since the derived $\Te$ values have been
greatly reduced, so are the model fluxes, and larger solid angles are
thus required to fit the photometric data, implying larger stellar
radii or smaller masses, as seen in Figure \ref{fg:f8}.
 
An interesting consequence of our findings concerns the mass
distribution. An intriguing question raised by BLR was the fact that
the mean mass of their subsample of cool, helium-atmosphere white
dwarfs, $\langle M\rangle=0.72$~\msun, was significantly higher than
that obtained for their likely progenitors, the DB stars (see Fig.~22
of BLR). Indeed \citet{beauchamp96} obtained a mean mass of $\langle M
\rangle=0.59$ \msun\ based on spectroscopic fits to the optical spectra of 
54 DB stars, in excellent agreement with the mean for DA stars derived
by \citet{BSL}, for instance\footnote{\citet{BSL} used thin
hydrogen evolutionary models to derive a mean mass for DA stars of
0.56 \msun, while the revised mean mass obtained by \citet{B95} based
on thick hydrogen models yields precisely 0.59 \msun.}. Figure
\ref{fg:f9} compares the mass distribution of our 16 DQ stars analyzed
with pure helium models (top panel) and models including carbon
(bottom panel) with the spectroscopic mass distribution of DB stars of
\citet{beauchamp96}. Our improved analysis brings the mean mass of DQ
stars from $\langle M\rangle=0.73$ \msun\ down to $\langle M
\rangle=0.62$ \msun, in much better agreement with the spectroscopic
mean mass of DB stars, and more in line with our understanding that DB stars
represent the mosty likely progenitors of DQ stars.

It is worth mentioning that trigonometric parallaxes are useful
specifically to derive masses for the stars in our sample. Had we
fixed the surface gravity to the canonical value of $\logg=8.0$, the
derived effective temperatures and carbon abundances would have
changed by at most 15 K and 0.05 dex, respectively, with respect to
the values given in Table 1. This is because the broadband fluxes and
the strength of the Swan bands are not particularly sensitive to
$\logg$ (see BRL and Fig.~\ref{fg:f2}).

\subsection{Analysis of the DQ White Dwarfs in the SDSS}

Prior to the SDSS study, only 37 cool ($\Te < 15,000$~K) white
dwarfs with carbon features had been analyzed by various
investigators, and the SDSS first data release {\it alone} has more
than doubled the number of DQ stars known, allowing us to improve
significantly the statistics of DQ stars. The method we use here to
fit the SDSS data is similar to that described above, with the
exception that the SDSS $ugriz$ photometry is used instead.  These
photometric passbands cover the entire optical range from the UV
atmospheric cutoff (3200~\AA) to the red sensitivity cutoff of the
detector ($\sim 10,000$~\AA). Furthermore, since trigonometric
parallax measurements are not available for the SDSS stars, we assume
a value of $\logg=8.0$ for all objects.

Figure \ref{fg:f10} shows an example of our best fit to the DQ
star SDSS J135628.25$-$000941.2 whose optical spectrum exhibits strong
molecular Swan bands. As before, we display our pure helium fit as
well as our solution with models including carbon. In this particular
case, the photometric fit is clearly better when carbon is
included. Our results for the DQ stars from the SDSS sample are given
in Table 2, and sample fits to the optical spectra are displayed in
Figure \ref{fg:f11}. We see that the most important C$_2$
features are well reproduced by our model spectra. Note that the
polynomial correction function used in our fitting procedure is
reduced to only the first term for these fits, which corresponds
simply to the scaling factor in equation (2).  This indicates (1) that
the the SDSS spectra are well calibrated spectrophotometrically, and
(2) that our predicted continuum slope in the optical is internally
consistent with our photometric temperature determination.

In some DQ stars shown in Figure \ref{fg:f11}, the
$\Delta\nu=+2$ bands near 4380 \AA\ are also visible, and in all cases,
they are predicted somewhat too deep. This was also noticed by
\citet{yackovich82} who proposed possible explanations for this
discrepancy. For instance, by neglecting the pressure-dissociation of
the C$_2$ molecule (analogous to the lowering of the ionization
potential in atomic systems), we diverge from the correct populations
as higher vibrational states are considered. Alternatively, the
molecular data for these transitions may be simply inaccurate.  Since
all the other bands are fairly well reproduced by our models, we feel
that our carbon abundances are not affected by this problem.

Finally, we note that the results presented here have been obtained
without applying any correction to the photometric measurements to
account for the extinction from the interstellar medium. Had
we applied the {\it full} correction obtained from the reddening maps
of \citet{schlegel98}, as given by the $A_g$ values taken from
\citet{kleinman04}, the derived atmospheric parameters would have 
differed by less than 500 K in $\Te$ and 0.5 dex in $\che$ in most
cases, but these differences would have reached as much as 1000 K and
1.5 dex, respectively, in some extreme cases. However, the photometric
distances for these objects given in Table 2, obtained under the
assumption of $\logg=8.0$, indicate that most objects are within 150
pc, and that none exceeds 300 pc. Consequently, they are relatively close
by, and only a small fraction of the absorption on that line of sight
should be applied. The corresponding changes in the atmospheric
parameters should therefore be small.

\subsection{Effects of Hydrogen and Heavy Elements}\label{metal}

All 56 DQ stars analyzed in this paper are featureless
near \halpha. Therefore the hydrogen abundance cannot be determined
from the optical spectroscopic observations, and only upper limits can
be set.  Limits on the hydrogen abundance in helium-atmosphere white
dwarfs depend sensitively on the range of effective temperature
considered. BRL obtained rough estimates of these limits by
calculating the hydrogen abundance at which a weak \halpha\ feature is
produced just above the detection threshold. Their results indicate
that the absence of \halpha\ implies $\nh<-2$ at $\Te=6500$~K and
$\nh<-4$ at 10,000~K. The energy distributions for such small hydrogen
abundances are indistinguishable from those obtained from pure helium
models, and the derived atmospheric parameters are not affected
significantly.

To explore further this issue, we re-fitted the DQ stars from the SDSS
sample with a model grid including a small trace of hydrogen of
$\nh=-5.0$. Our results reveal that \halpha\ is not visible at any effective
temperature, and that the derived atmospheric parameters remain
unchanged. The same exercise with a larger hydrogen abundance of
$\nh=-3.0$ produces a weak \halpha\ feature for stars above
$\Te\sim7000$~K. Such high hydrogen abundances are of course ruled out
by our spectroscopic observations. For cooler stars, however, 
\halpha\ remains undetected at this hydrogen abundance, and
a fair amount of hydrogen could thus be hidden in the atmosphere of
cool DQ stars. Our fits with $\nh=-3.0$ yield effective temperatures
hotter by up to 200 K and carbon abundances increased by up to 0.6 dex
when compared to the solutions derived from hydrogen-free
models. It is even possible that more hydrogen is hidden in these
objects, but more work is required to properly assess this
problem. We therefore conclude that it is safe to assume hydrogen-free
models to analyze DQ stars above 7000 K, while atmospheric parameters
for cooler stars are more uncertain due to the unknown amount of
hydrogen present in their atmospheres.

Various studies have already assessed the importance of heavy elements
in the calculations of cool DQ models
\citep{grenfell74,koester82,wegner84}.  Their results can be summarized
as follows. As long as the principal electron donor remains carbon,
traces of other metals are not important. Given that there is no trace
of metallic lines in our optical spectra, nor any molecular bands of
CO and CN, and given that upper limits compatible with UV observations
($IUE$) constrain the abundance of heavy elements to relatively low
values \citep{koester82}, the presence of heavy elements should not
affect the thermodynamic structures and atmospheric parameters derived
from our mixed helium and carbon models.

\section{RESULTS}\label{results}

\subsection{Thickness of the Helium Layer in DQ Stars}

Our results are summarized in Figure \ref{fg:f12} where the carbon
abundances for all objects in both the BRL/BLR and SDSS samples are
shown as a function of effective temperature. We see that the DQ stars
form a relatively narrow sequence in the $\che-\Te$ plane; we ignore
for the moment the few objects that form a parallel sequence about 1
dex above the bulk of DQ stars (see \S~\ref{massive} below). We
attribute this rather remarkable finding to the homogeneity of our
approach. The sequence defined by the SDSS sample overlaps the BRL/BLR
sequence quite nicely, suggesting that the physical characteristics of
these two samples are very similar. Since the BRL/BLR sample as a
whole has a normal mean mass near 0.6 \msun, the value of $\logg=8$
assumed for the SDSS sample is probably reasonable. It is also
comforting to see that Procyon B (star symbol at 7740 K), with a
precise mass determination of 0.602
\msun\ \citep{provencal02}, falls well within our sequence.

Also shown in Figure \ref{fg:f12} are the results from two
different sets of evolutionary model calculations, which predict the
carbon abundance as a function of decreasing effective temperature for
various values of the thickness of the helium layer $q({\rm He})$. The
dotted curves represent the calculations of
\citet{pelletier86} at 0.6 \msun\ with, from top to bottom,
$\log q({\rm He})=-4.0$, $-3.5$, and $-3.0$. According to these
models, our abundance determinations would suggest values of $\log
q({\rm He})$ smaller than $-3.5$ for most DQ stars, much thinner than
the values predicted by post-AGB evolutionary models, as discussed in
the Introduction. A new generation of evolutionary models computed by
\citet{fon05} yield quite
different results, however, as shown in Figure \ref{fg:f12} by
the solid curves for models at 0.6 \msun\ with, from top to bottom,
$\log q({\rm He})=-4.0$, $-3.0$, and $-2.0$. With these new models,
the carbon abundances in DQ stars are consistent with values of the
helium layer thickness between $\log q({\rm He})=-3$ and $-2$, in much
better agreement with the results from post-AGB models. We do not show
here the predictions of the other existing set of theoretical
calculations of the carbon abundance in DQ stars, that of
\citet{mhj98}, since they suggest a monotonic $increase$ of the carbon
abundance with decreasing effective temperature (see their Figs.~4
and 5), in clear conflict with the very existence of the sequence
uncovered in the present work.

We note that the models of \citet{fon05}, while still preliminary,
incorporate significant improvements over those constructed by
\citet{pelletier86}. Firstly, they result from an evolutionary code
that fully takes into account the feedback effects of diffusion on the
evolving models, in contrast to the semi-evolutionary approach of
\citet{pelletier86}. Secondly, they include improved constitutive
physics at the level of the envelope equation of state and of the
radiative opacity \citep[see, e.g.,][]{fon01}. In particular, the
use of the OPAL opacity data to replace the older Los Alamos opacity
tables used in Pelletier et al.~leads to deeper convection zones than
before. This fact alone explains in large part the seemingly very
different results shown in Figure \ref{fg:f12} between the two generations of
model calculations (for more on this, see \citealt{fw97} and
\citealt{mhj98}). Thirdly, the basic paradigm has changed since
the Pelletier et al.~era when it was assumed that helium had already
separated from carbon in the hot phases of the evolution, whereas models
of PG~1159 stars are used as feeder models in the newer calculations and
the helium envelope builds up with passing time as the result of carbon
and oxygen settling. Thus, in the more recent picture, the carbon
pollution is not produced by carbon migrating from the core, but by the
development of a helium superficial convection zone whose base catches
up some of the settling carbon and dredges it up to the surface. 

We further note that there is much to be explored in the \citet{fon05}
approach, including a discussion of the effects of changing various
parameters such as the total mass, the convective efficiency, the
amount of turbulence, and the initial PG~1159 envelope abundances. In
addition, the computations need to be pushed to lower effective
temperatures than has been done in this paper. In this connection, it
is the unavailability of OPAL data for carbon at low temperatures that
currently prevents the extension of the calculations to lower
effective temperatures than those shown in Figure \ref{fg:f12} for the
three solid curves. This and other problems will need to be addressed
elsewhere. For the purposes of the present paper, the newer
calculations available are sufficient to indicate that a helium layer
of expected thickness has built up in the bulk of the DQ white dwarfs.

We have also calculated the abundances at which the carbon features
disappear, as a function of effective temperature. This is shown by
the dashed curve in Figure \ref{fg:f12}, which represents the
detection threshold of the C$_2$ Swan bands below $\Te=10,000$~K, or
of the atomic C~\textsc{i} lines above 10,000~K. These results reveal
that cool, helium-rich atmosphere white dwarfs with helium envelopes
thicker than $\log q({\rm He})\sim-2.5$ would appear as featureless DC
stars (or DZ stars if metals are present). Thus, DQ stars can be
interpreted as white dwarfs with thin enough helium envelopes for
their atmospheres to be polluted by settling carbon brought back to the
surface through convective dredge-up. 
We note also that the progenitors of the bulk of the DQ stars analyzed
here --- hotter DB stars with comparable thicknesses of the helium
layer --- are not expected to show neutral carbon lines in their
optical spectra since the predicted abundances for $\log q({\rm
He})\geq-3$ fall below the detection threshold shown in Figure
\ref{fg:f12}.

\subsection{Massive DQ Stars}\label{massive}

As mentioned above, several objects in Figure \ref{fg:f12} show
carbon abundances that lie about 1 dex above the bulk of DQ
stars. A possible explanation would be to consider these objects as
average-mass stars with thinner than average helium layer envelope,
i.e., with $\log q({\rm He})$ between $-4.0$ and $-3.0$ or so. However,
the dichotomy of the distribution shown in Figure \ref{fg:f12} argues against that
interpretation as we should presumably observe a continuun of values of
$\log q({\rm He})$ instead of two apparently well separated sequences. 
Another interpretation rests on the parallax value of one of the stars
which seem to populate the sequence of more heavily polluted DQ
stars. And indeed, among these, the only white dwarf with a measured
trigonometric parallax, and thus with a mass determination, is G47-18
(WD~0856+331; filled circle at the top) with an estimated mass of
$M=1.05$ \msun, by far the most massive DQ star in Table 1 (see also the
mass distribution in Fig.~\ref{fg:f9}). It is the only massive DQ
star in the trigonometric parallax sample, the other objects being
concentrated between the $\log q({\rm He})=-3$ and the $\log q({\rm 
He})=-2$ sequences of \citet{fon05} and their extrapolations at low 
$\Te$. It is thus tantalizing to associate the DQ stars
with larger-than-average carbon abundances with the high mass
component of the white dwarf mass distribution. \citet{liebert03} has
indeed suggested that the hot DQ stars discovered in the SDSS could
all be massive, and could correspond to the missing high mass tail of
the DB star mass distribution determined by
\citet{beauchamp96}. If both hypotheses are confirmed, the cool DQ
stars with large carbon abundances identified here would simply
represent the natural extension of the hot DQ stars identified by
\citet{liebert03}. Similarly, the unique DQ white dwarf discovered
by \citet{carollo02} and shown in Figure \ref{fg:f12} (star
symbol at 5120 K) would also be massive.

Finally, our results also explain why the rare DQ stars showing atomic
lines in their optical spectra (G47-18, G227-5 and G35-26) have
estimated masses above $\sim 1$ \msun\ \citep{liebert03} since, as
explained in the previous section, stars with average helium layer
thicknesses --- and a mean mass near $\sim$0.6 \msun\ --- are not
expected to show detectable carbon lines. We also note that the  
four hottest DQ stars in our SDSS sample all show atomic carbon lines
and have high carbon abundances. It is indeed expected that high carbon
pollution in massive stars must be the result of relatively thin helium
envelopes \citep{pelletier86}, but it remains to be seen if modern 
evolution/diffusion calculations extended to high mass models can
account quantitatively for this interpretation.

\subsection{Temperature Distribution of DQ Stars}

The temperature distribution of the DQ stars in the BRL/BLR and SDSS
samples is displayed in Figure \ref{fg:f13}. We first observe that
between $\Te\sim 10,000$~K and 8000~K, the number of DQ stars
increases steadily. This is most easily explained in terms of the
limit of visibility of the carbon features. Indeed, the separation in
Figure \ref{fg:f12} between the DQ empirical sequence and the
detection threshold (dashed line) is shown to {\it increase} with
decreasing effective temperature, making the DQ phenomenon easier to
detect at low temperatures, at least in the optical. The only DQ stars
that could be identified at high effective temperatures are the
massive ones according to our previous interpretation, and there are
usually few massive white dwarfs in magnitude-limited samples.

On the cool side of the temperature distribution, we observe a sudden
drop in the number of DQ stars below $\Te\sim 7000$~K,
thanks to the SDSS sample, which makes this result more statistically
significant. At $\Te=6000$~K, there is even an abrupt cutoff below
which only one object in our sample of 56 DQ stars is found.
This number increases to two if we include the DQ star reported by
\citet{carollo02}. However, if our interpretation is correct, both 
of these cool stars are also massive according to the results of
Figure \ref{fg:f12}. Hence, it is possible that whatever
mechanism is responsible for the temperature distribution cutoff of
normal mass DQ stars, more massive DQ stars may persist at lower
effective temperatures. Note that this cutoff was also noticed by BRL,
although their sample was much smaller (see
Fig.~\ref{fg:f13}). Finally, we mention that a visual extrapolation of
the observed sequence of normal DQ stars in Figure \ref{fg:f12}
suggests carbon abundances near $\che\sim-8.5$ for temperatures near
$\Te\sim5000$~K. DQ stars with such low carbon abundances could still be
easily detected in this temperature range according to the results
shown in Figure \ref{fg:f2}, reinforcing our conclusions that
the observed cutoff in the DQ temperature distribution is real.

Can this be a selection effect? As discussed by \citet{kleinman04},
the principal science objective of the SDSS was to obtain redshifts of
distant galaxies and QSO. Potential targets were selected for
spectroscopic follow-up on the basis of the five-band $ugriz$
photometry. The different targeting categories, with different
criteria and priorities, are used to assign all the fibers available
on the spectroscopic plate. If there are not enough primary targets to
fill a given plate, lower targeting categories are assigned
fibers. Since the white dwarf targeting category has a low priority,
color and proper motion information are used to find candidates in
other SDSS targeting categories in order to have a more complete
sample of white dwarf candidates. These candidates were then manually
inspected and sorted in various white dwarf spectral types. According
to our calculations, cool ($\Te<6000$~K) DQ white dwarfs are expected
to have $u-g\sim0.7-1.0$ and $g-r\sim 0.5-0.8$ and should have been
easily detected according to the color criteria given in Table~2 of
\citet{kleinman04}. This is illustrated in Figure \ref{fg:f14} where we show
a $u-g$ vs.~$g-r$ color-color diagram similar to that displayed in
Figure 1 of \citet{harris03}. Small dots in this figure indicate
objects with stellar images while filled triangles represent the DQ
stars from Table 2. Superposed on this figure are our DQ photometric
sequences at $\logg=8$ for various values of $\Te$ and $\che$. Note
that the three SDSS stars in the 6000-6500 K bin in Figure
\ref{fg:f13} actually have effective temperatures very close to 6500~K
(see Table 2), in agreement with the results shown here. DQ stars
significantly cooler than $\Te=6500$~K should have been detected if
they existed according to the results shown in Figure \ref{fg:f14}. We note
finally that the absence of extremely cool DQ stars is also observed
in the BRL/BLR data, a sample based only on common proper motion
criteria.

The physical explanation for the existence of this temperature cutoff
remains speculative at this point. But as discussed in the
Introduction, one possibility is that the C$_2$ features give place to
C$_2$H features below $\sim6000$~K \citep{schmidt95}. For C$_2$H to
become one of the dominant molecules in the photosphere of a
helium-dominated white dwarf atmosphere, large amounts of
hydrogen of the order of $\nh\sim -1$ are required. Some mechanisms
responsible for the sudden enrichment of DQ stars with hydrogen have
been proposed by BRL (see the discussion in their \S~6.3.2). Several
of these so-called C$_2$H stars have been discovered in the SDSS
survey \citep[see, e.g., Fig.~8 of][]{harris03}. They cannot be analyzed
within our current theoretical framework, however, since detailed
theoretical absorption profiles do not exist yet for C$_2$H bands in
the optical. The existence of many DQ stars above $\Te\sim6000$~K and
the existence of C$_2$H stars below this temperature appear more than
coincidental, and C$_2$H stars could well hold the key to our
understanding of the nature of the cutoff in the temperature
distribution of DQ stars.

\section{SUMMARY AND CONCLUSIONS}\label{conclusion}

A detailed photometric and spectroscopic analysis of 56 DQ white
dwarfs was presented. Improved model atmosphere calculations including
atomic and molecular carbon were used to determine effective
temperatures and carbon abundances for two different samples of DQ
stars (16 stars in the BRL/BLR sample and 40 stars in the SDSS
sample). This represents the largest set of DQ white dwarfs analyzed
in a {\it homogeneous} fashion. A summary of our findings follows:

\begin{enumerate}

\item Our reanalysis of the DQ stars in the BRL/BLR sample indicates that 
effective temperatures derived from model atmospheres including carbon
are significantly lower than the temperatures obtained from pure helium models,
in agreement with the conclusions of \citet{provencal02} for Procyon
B. The explanation of this effect is that the additional free
electrons provided by carbon increases the He$^-$ free-free opacity,
which in turn affects the atmospheric structure in the continuum
forming region.

\item Our sample contains 16 stars with trigonometric parallax measurements
from which stellar masses can be obtained. The mean mass of this
sample derived from pure helium models is $\langle M \rangle$= 0.73
\msun, significantly higher than the spectroscopic mean mass of DB stars, 
$\langle M \rangle$= 0.59 \msun. We now find with our models including
carbon that the mean mass of the DQ star sample is reduced to $\langle
M \rangle=0.62$ \msun, in much better agreement with the spectroscopic
mean mass of DB stars, and more in line with our understanding that DB
stars represent the most likely progenitors of DQ white dwarfs.

\item We determined effective temperatures and carbon abundances for 
40 new DQ stars discovered in the first data release of the Sloan
Digital Sky Survey. This more than doubles the number of DQ stars ever
analyzed.

\item We find that DQ stars form a well defined sequence in a 
$\che$ versus $\Te$ diagram, with the carbon abundance decreasing
monotonically with decreasing effective temperature. The evolutionary
models of \citet{pelletier86} overplotted in this diagram suggest that 
the thickness of the helium layer is around $\log q({\rm He}) \sim
-3.5$ for the bulk of DQ stars, a value smaller than the predictions
from post-AGB models by more than an order of magnitudes. Improved
evolutionary models by \citet{fon05} brings the inferred thickness
of the helium layer to a value between $\log q({\rm He})= -3$ and
$-2$, more compatible with post-AGB theory.

\item Several DQ stars in our sample show larger than average carbon 
abundances. Among these, only G47-18 has a mass determination, and it
also turns out to be the most massive object in our sample. Since all
other DQ stars in our trigonometric parallax sample have normal carbon
abundances and inferred stellar masses close to 0.6 \msun, we argue
that the DQ stars with larger than average carbon abundances are
massive, and could represent the high mass tail of the white dwarf
mass distribution. Their hotter counterpart are probably the hot DQ
stars reported by \citet{liebert03}, who suggested that these were all
massive, and could represent the missing high mass tail of the DB star
spectroscopic mass distribution.

\item Finally, the number distribution of DQ white dwarfs as a function of 
temperature clearly shows a sudden drop at about $\Te\sim$7000 K and
what seems to be an abrupt cutoff at $\Te\sim$6000 K (or even
above). This cutoff, also noticed by BRL, appears now statistically
more significant with the addition of the SDSS stars. The physical
mechanism responsible for this cutoff is still unknown even though it
is believed to be somehow related to the existence of C$_2$H stars
(see BRL and BLR).

\end{enumerate}

Future work on DQ white dwarfs should eventually include archival
spectra of stars showing carbon in the ultraviolet ($IUE$ and $HST$),
as well as an updated analysis of the SDSS sample when all the data
are released.  Further studies, which are underway, will also include
the DZ stars from the BRL/BLR and SDSS samples, from which we hope to
draw a clearer picture of the chemical evolution of cool white dwarfs.

\acknowledgements{We wish to thank S.~Kleinman for useful discussions 
concerning the SDSS data, and A.~Gianninas and C.~Pereira for some of
the spectroscopic data acquisition. We also wish to thank M.~T.~Ruiz
and the late C.~Anguita for providing us with their unpublished
trigonometric parallax measurements. This work was supported in part
by the NSERC Canada and by the FQRNT (Qu\'ebec).}

\clearpage
 \clearpage
 \begin{deluxetable}{llcccc}
 \tabletypesize{\footnotesize}
 \tablecolumns{5}
 \tablewidth{0pt}
 \tablecaption{Atmospheric Parameters for DQ Stars from the BRL/BLR Sample}
 \tablehead{
 \colhead{WD} &
 \colhead{Name} &
 \colhead{$T_{\rm eff}$(K)} &
 \colhead{log $g$} &
 \colhead{$M/M_{\odot}$}&
 \colhead{log (C/He)}}
 \startdata
0115$+$159 &LHS 1227      &  9050 (310)& 8.19 (0.07)& 0.69 (0.04)& $-$4.33 (0.32)\\
0341$+$182 &Wolf 219      &  6510 (130)& 7.99 (0.10)& 0.57 (0.06)& $-$6.41 (0.10)\\
0435$-$088 &L879-14       &  6300 (110)& 7.93 (0.04)& 0.53 (0.02)& $-$6.41 (0.06)\\
0548$-$001 &G99-37        &  6070 (100)& 8.18 (0.04)& 0.69 (0.03)& $-$6.82 (0.06)\\
0706$+$377 &G87-29        &  6590 (140)& 7.98 (0.10)& 0.56 (0.06)& $-$6.40 (0.12)\\
 \\
0856$+$331 &G47-18        &  9920 (240)& 8.74 (0.08)& 1.05 (0.05)& $-$2.88 (0.08)\\
0935$-$371A&LDS 275A      &  9380 (390)& 8.32 (0.06)& 0.78 (0.04)& $-$4.21 (0.42)\\
0946$+$534 &G195-42       &  8100 (240)& 8.27 (0.12)& 0.75 (0.08)& $-$5.33 (0.36)\\
1115$-$029 &LHS 2392      &  9270 (420)& 7.90 (0.36)& 0.52 (0.17)& $-$4.25 (0.44)\\
1142$-$645 &LHS 43        &  7900 (220)& 8.07 (0.02)& 0.62 (0.01)& $-$5.14 (0.20)\\
 \\
1157$-$462 &ESO 267-110   &  7190 (250)& 8.32 (0.15)& 0.78 (0.10)& $-$5.66 (0.12)\\
1831$+$197 &G184-12       &  7110 (170)& 7.38 (0.63)& 0.28 (0.16)& $-$5.87 (0.16)\\
2059$+$316 &G187-15       &  9360 (350)& 7.87 (0.22)& 0.50 (0.11)& $-$3.99 (0.32)\\
2140$+$207 &LHS 3703      &  8200 (250)& 7.84 (0.06)& 0.49 (0.04)& $-$5.28 (0.42)\\
2311$-$068 &G157-34       &  7440 (190)& 8.09 (0.20)& 0.63 (0.12)& $-$5.67 (0.20)\\
 \\
2352$+$401 &G171-27       &  7710 (200)& 7.78 (0.28)& 0.45 (0.12)& $-$5.19 (0.16)\\
 \enddata
 \end{deluxetable}
 \clearpage

\clearpage
 \clearpage
 \begin{deluxetable}{lccrccc}
 \tabletypesize{\footnotesize}
 \tablecolumns{6}
 \tablewidth{0pt}
 \tablecaption{Atmospheric Parameters for DQ Stars from the SDSS}
 \tablehead{
 \colhead{Name} &
 \colhead{Plate} &
 \colhead{MJD} &
 \colhead{Fiber} &
 \colhead{$T_{\rm eff}$(K)} &
 \colhead{log (C/He)}&
 \colhead{$D$ (pc)}}
 \startdata
SDSS J000011.57$-$085008.4 &  650&52143&  450&  7800 (200)&$-$5.29 (0.22)& 153\\
SDSS J000052.44$-$002610.5 &  387&51791&  168&  7210 (140)&$-$6.16 (0.30)& 112\\
SDSS J000807.54$-$103405.6 &  651&52141&  199&  7600 (190)&$-$5.53 (0.24)& 139\\
SDSS J002531.50$-$110800.9 &  653&52145&   86&  8540 (120)&$-$4.53 (0.22)& 107\\
SDSS J014619.96$-$082616.9 &  664&52174&  402&  7240 (150)&$-$6.13 (0.32)&  70\\
 \\
SDSS J015433.57$-$004047.2 &  403&51871&  268&  7360 ( 80)&$-$5.70 (0.20)& 112\\
SDSS J015441.74$+$140308.0 &  430&51877&  558&  6450 ( 70)&$-$6.97 (0.22)&  59\\
SDSS J020906.07$+$142520.8 &  428&51883&  364&  7140 (120)&$-$5.87 (0.18)& 139\\
SDSS J024332.74$+$010112.4 &  408&51821&  612&  8330 (190)&$-$3.77 (0.08)& 294\\
SDSS J032054.11$-$071625.4 &  460&51924&  236&  6570 ( 80)&$-$4.88 (0.20)& 129\\
 \\
SDSS J033218.22$-$003722.1 &  415&51810&  240&  8110 (150)&$-$4.88 (0.20)& 115\\
SDSS J035222.86$-$060506.3 &  463&51908&  559&  7510 (150)&$-$5.81 (0.30)& 120\\
SDSS J080843.15$+$464028.7 &  438&51884&   63&  5140 ( 30)&$-$6.66 (0.14)&  89\\
SDSS J083637.79$+$481752.5 &  550&51959&  433&  7250 (140)&$-$6.08 (0.28)& 115\\
SDSS J090200.36$+$503723.0 &  551&51993&  612&  7690 (160)&$-$5.43 (0.24)& 158\\
 \\
SDSS J091922.18$+$023605.0 &  473&51929&  458& 10670 (310)&$-$3.27 (0.30)& 237\\
SDSS J094004.64$+$021022.6 &  477&52026&  493&  7140 ( 90)&$-$5.96 (0.20)&  55\\
SDSS J094520.78$+$555838.0 &  556&51991&  571&  7350 (160)&$-$5.77 (0.22)& 182\\
SDSS J095137.60$+$624348.7 &  487&51943&  227&  8410 (220)&$-$4.72 (0.24)& 157\\
SDSS J101219.90$+$004019.7 &  502&51957&   95&  9180 (100)&$-$4.18 (0.34)& 107\\
 \\
SDSS J104346.74$+$030318.5 &  506&52022&  454&  6750 (110)&$-$6.31 (0.16)& 129\\
SDSS J113359.94$+$633113.2 &  597&52059&  139& 10980 (430)&$-$3.18 (0.36)& 290\\
SDSS J114851.68$-$012612.8 &  329&52056&  578&  9450 (120)&$-$3.25 (0.10)&  93\\
SDSS J125359.61$+$013925.6 &  523&52026&  252&  8350 (220)&$-$4.88 (0.30)& 170\\
SDSS J135134.44$+$662314.3 &  497&51989&  553&  8890 ( 80)&$-$4.51 (0.38)& 100\\
 \\
SDSS J135628.25$-$000941.2 &  301&51942&  231&  6510 ( 60)&$-$5.79 (0.08)&  81\\
SDSS J140625.70$+$020447.0 &  532&51993&  475&  7510 (130)&$-$5.53 (0.20)& 122\\
SDSS J140632.42$+$014838.3 &  532&51993&  234&  7670 (150)&$-$5.63 (0.30)& 108\\
SDSS J144407.25$+$043446.8 &  587&52026&  418&  9800 (270)&$-$2.98 (0.10)& 275\\
SDSS J144808.07$-$004755.9 &  308&51662&  145&  7000 ( 90)&$-$6.18 (0.20)&  90\\
 \\
 \\
SDSS J154810.66$+$562647.7 &  617&52072&  551&  7970 (150)&$-$5.09 (0.22)& 154\\
SDSS J155413.53$+$033634.5 &  595&52023&  373&  6480 ( 60)&$-$6.84 (0.20)&  96\\
SDSS J161315.37$+$511608.4 &  623&52051&  201&  7900 (150)&$-$5.24 (0.24)& 164\\
SDSS J164328.54$+$400204.3 &  630&52050&  386&  7250 (160)&$-$5.86 (0.20)& 148\\
SDSS J165538.51$+$372247.1 &  632&52071&   92&  8890 (110)&$-$4.37 (0.30)& 144\\
 \\
SDSS J172502.66$+$533842.2 &  359&51821&  166&  6500 (170)&$-$6.53 (0.14)& 188\\
SDSS J204624.45$-$071519.1 &  635&52145&  228&  7820 (150)&$-$5.48 (0.32)& 141\\
SDSS J205316.34$-$070204.3 &  636&52176&  267&  6390 ( 40)&$-$5.25 (0.14)&  96\\
SDSS J231030.26$-$005745.8 &  381&51811&   93&  7880 ( 90)&$-$4.06 (0.08)& 151\\
SDSS J234132.83$-$010104.5 &  385&51877&  126&  8300 (140)&$-$4.91 (0.28)& 116\\
 \\
 \enddata
 \end{deluxetable}
 \clearpage

\clearpage

\figcaption[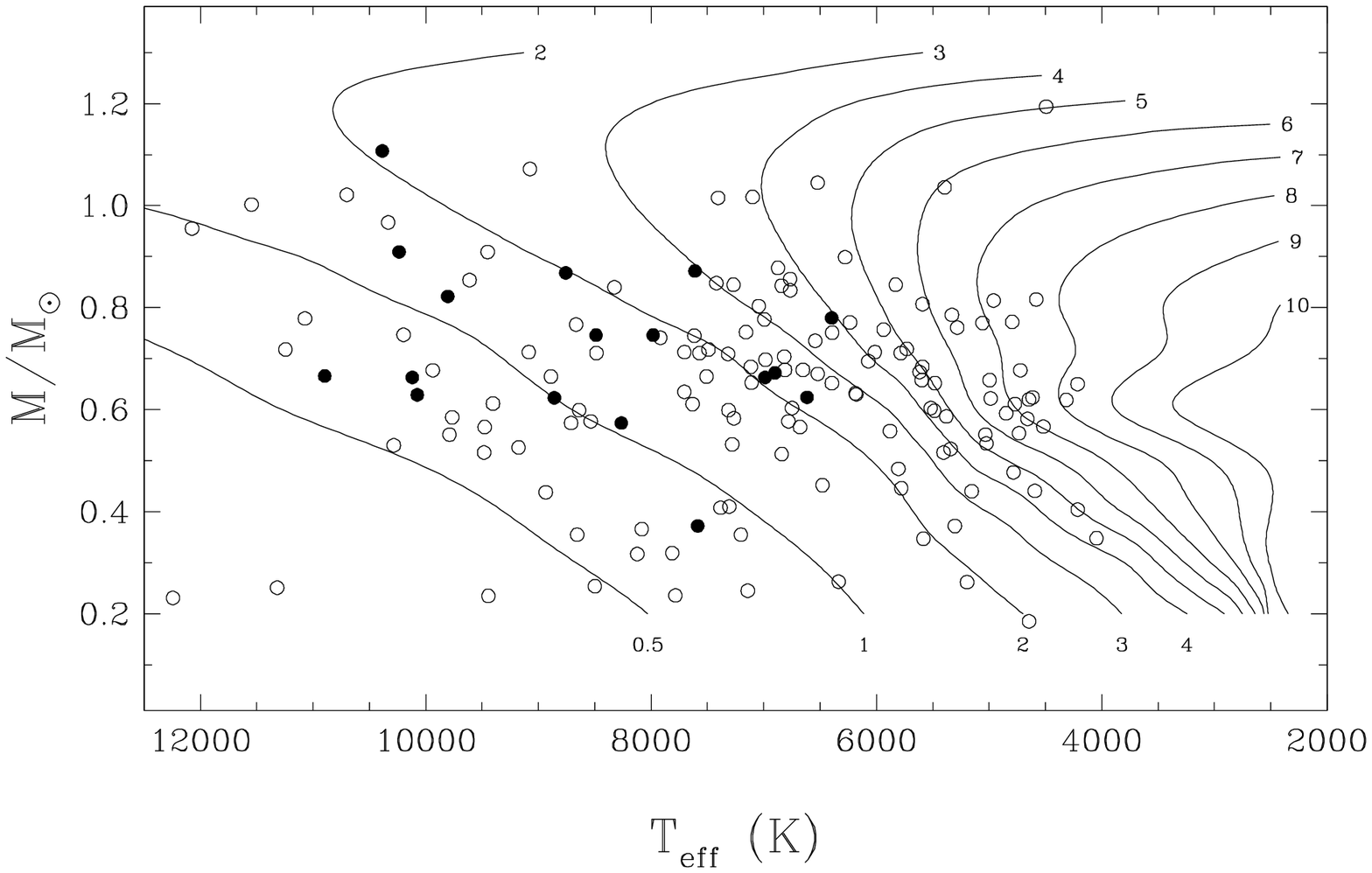] {Masses of white dwarfs in the
BLR trigonometric parallax sample as a function of effective
temperature.  DQ stars are indicated by filled circles. The stellar
parameters for these DQ stars --- obtained under the assumption of
pure helium compositions --- are taken directly from BLR, with the
exception of LDS 275A and ESO 267-110, which have been analyzed here
using new trigonometric parallax measurements reported in
\S~\ref{observation}. Also shown are
isochrones labeled in units of Gyr obtained from cooling sequences
with C/O core compositions, $q({\rm He})=10^{-2}$, and $q({\rm
H})=10^{-10}$. \label{fg:f1}}

\figcaption[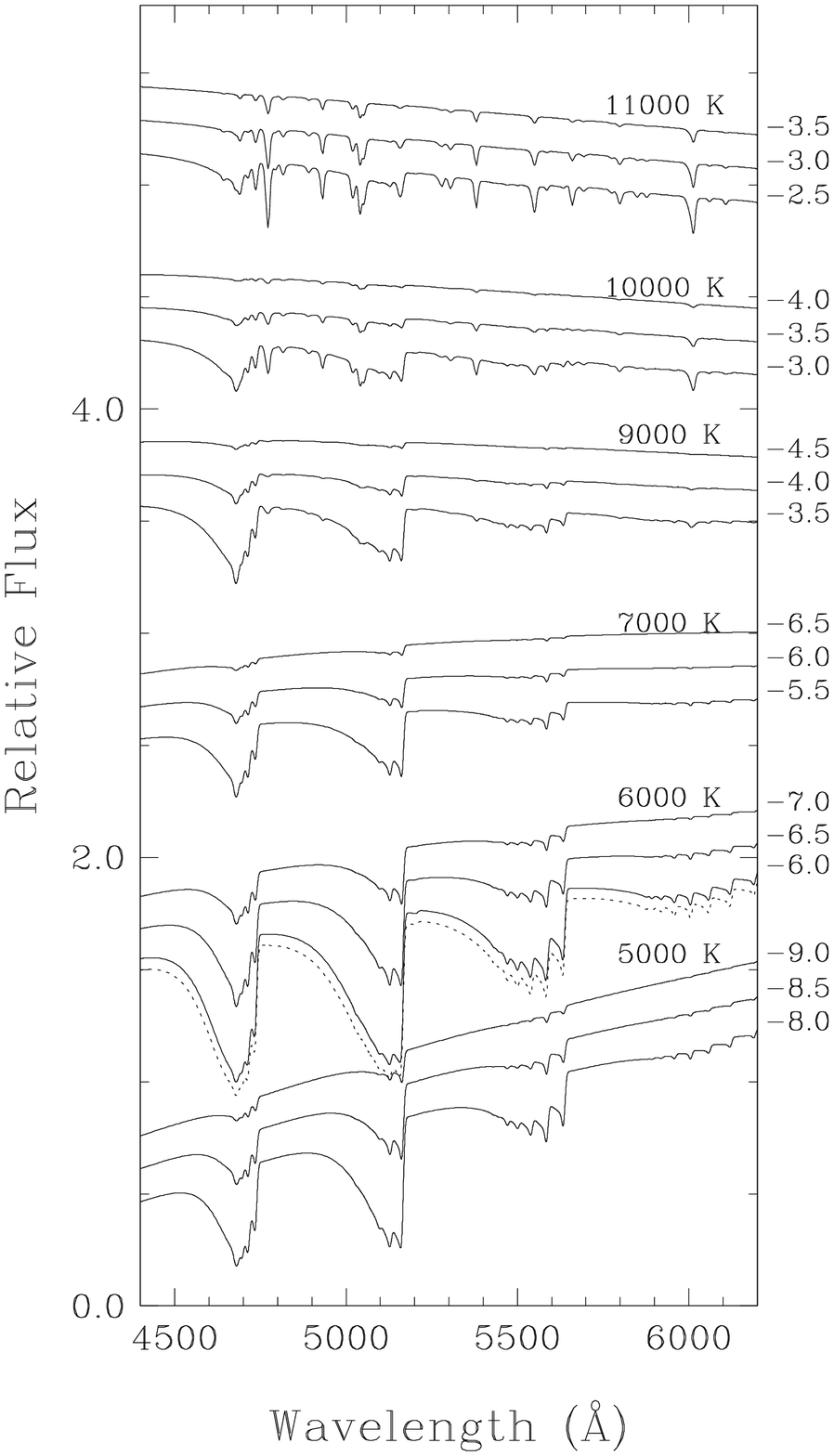] {Representative synthetic spectra of 
DQ white dwarfs taken from our model grid at $\logg=8.0$ and for various
carbon abundances. The spectra are normalized
to unity at $5200$ \AA\ and offset by an arbitrary factor for 
clarity. The dotted line corresponds to a $\Te=6000$~K, $\che=-6.0$ synthetic 
spectrum at $\logg=8.5$.\label{fg:f2}}

\figcaption[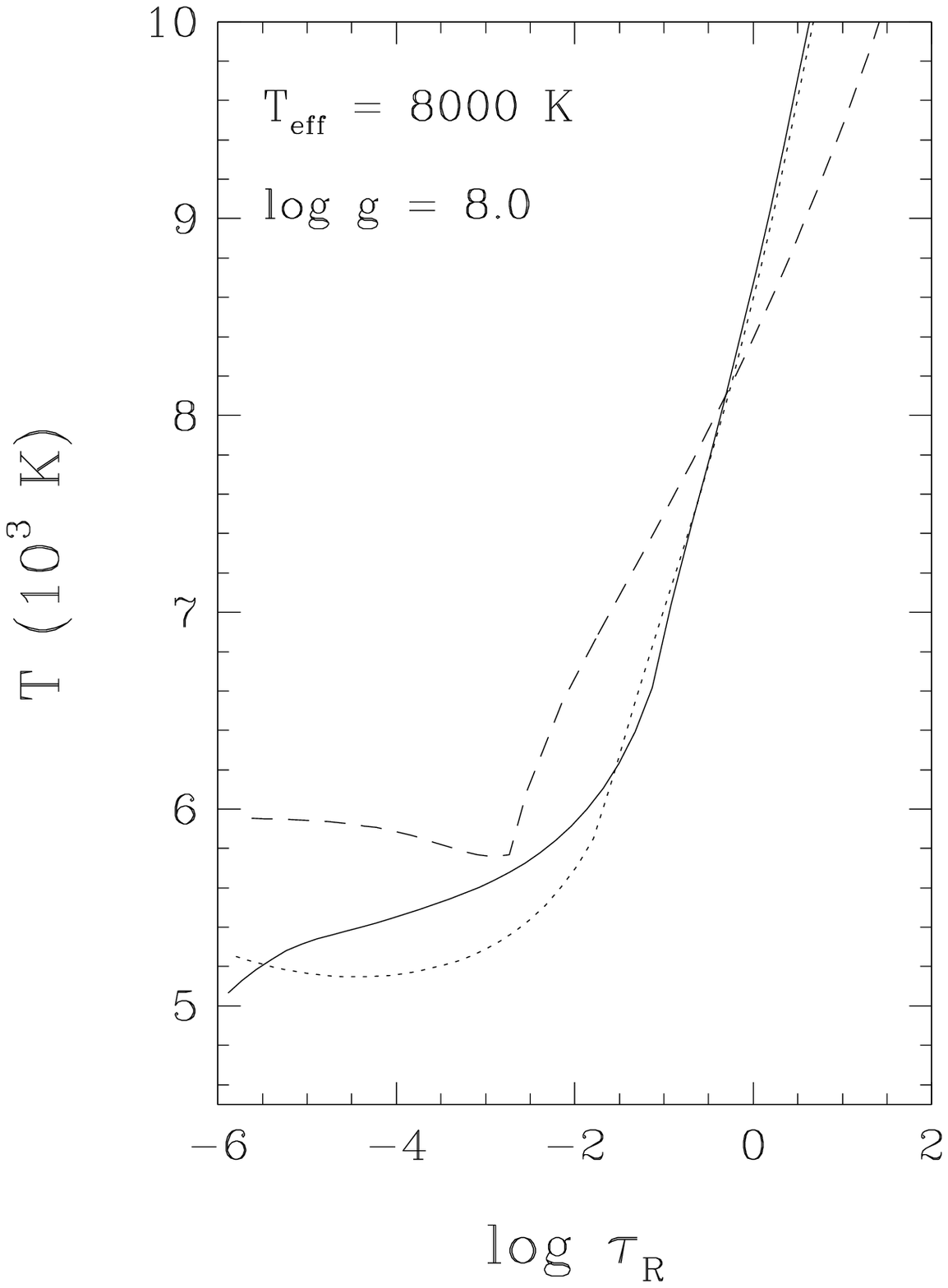] {Temperature as a function of Rosseland 
optical depth for white dwarf model atmospheres with $\Te=8000$~K and
$\logg=8$ calculated under the assumption of a pure helium composition
({\it dashed line}), a carbon abundance of $\che=-5.0$ included {\it
only} in the equation-of-state calculation ({\it dotted line}), and
the same carbon abundance but with the blanketing by carbon lines
and molecular bands taken into account ({\it solid
line}).\label{fg:f3}}

\figcaption[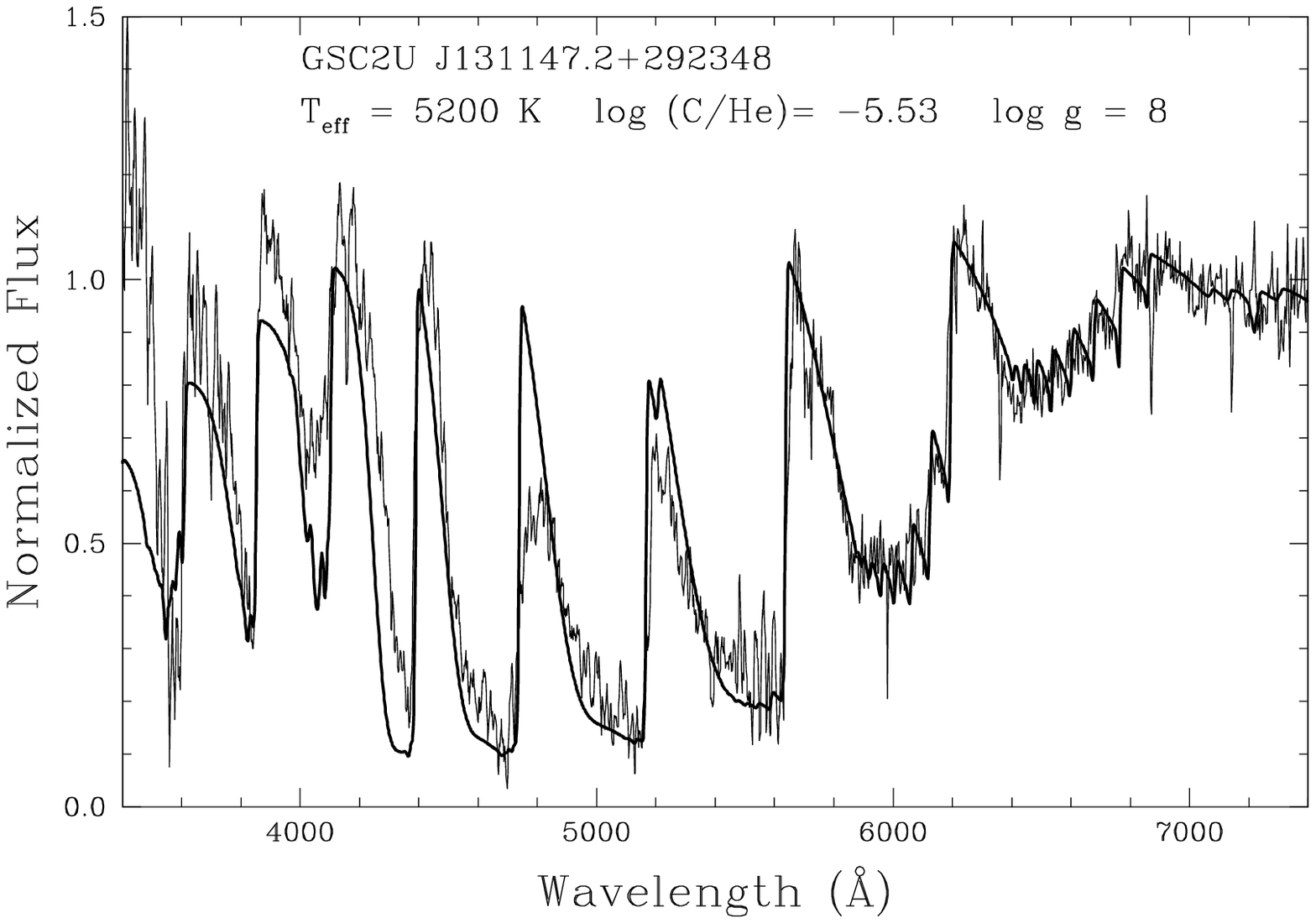] {Comparison of the optical spectrum 
of GSC2U J131147.2+292348 taken from \citet{carollo02} with a
synthetic spectrum interpolated from our model grid at the atmospheric
parameter solution obtained by
\citet{carollo03} and given here in the figure.\label{fg:f4}}

\figcaption[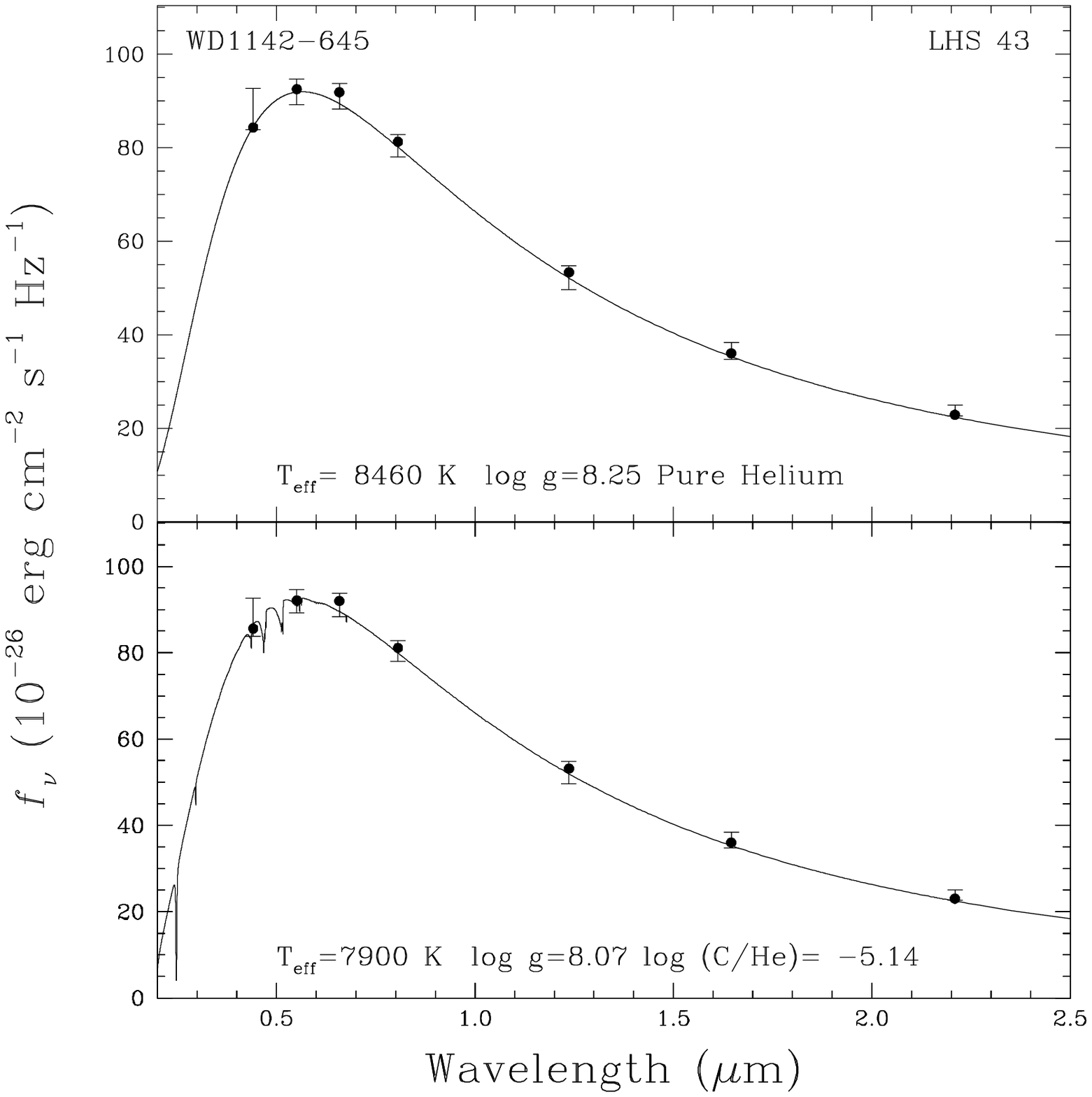] {Fits to the energy distribution
of LHS 43. The $BVRI$ and $JHK$ photometric observations are
represented by errors bars, while the filled circles correspond to the
average model fluxes used in the $\chi^2$ fit. The top and bottom
panels represent the pure helium solution and the solution including
carbon, respectively. The atmospheric parameters are given in each
panel. Also shown are the corresponding monochromatic Eddington fluxes
({\it solid lines}).\label{fg:f5}}

\figcaption[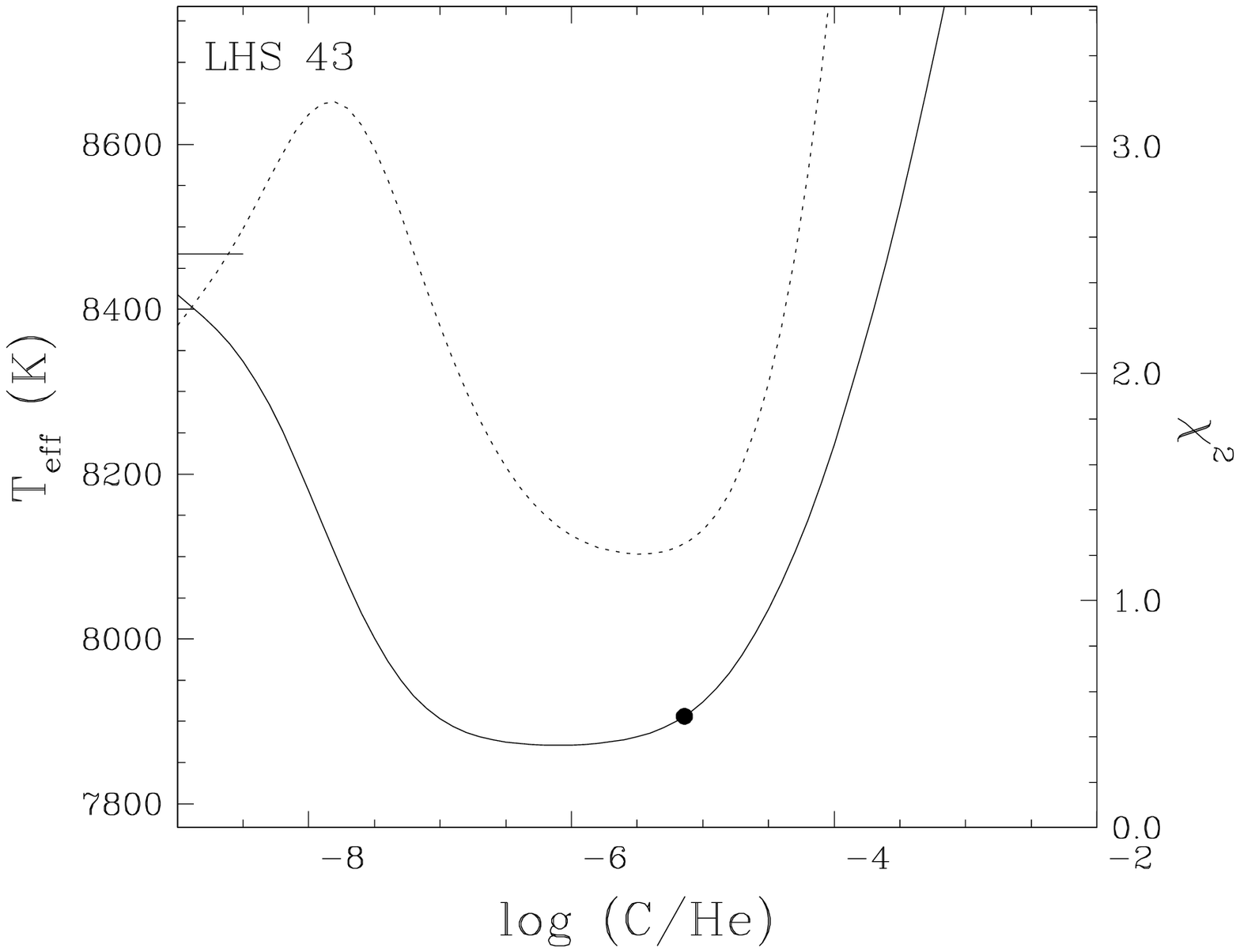] {{\it Solid line:} Effective temperature 
solution for LHS 43 based on fits to the energy distribution as a
function of the assumed carbon abundance; the solid dot represents the
final solution obtained from our iterative spectroscopic and
photometric scheme, $\Te=7900$~K, $\logg=8.07$, and $\che=-5.14$. {\it
Dotted line:} Value of the $\chi^2$ associated with the photometric
fit. The pure helium solution at $\Te=8460$~K and $\logg=8.25$ is
shown on the left hand side by the tick mark.\label{fg:f6}}

\figcaption[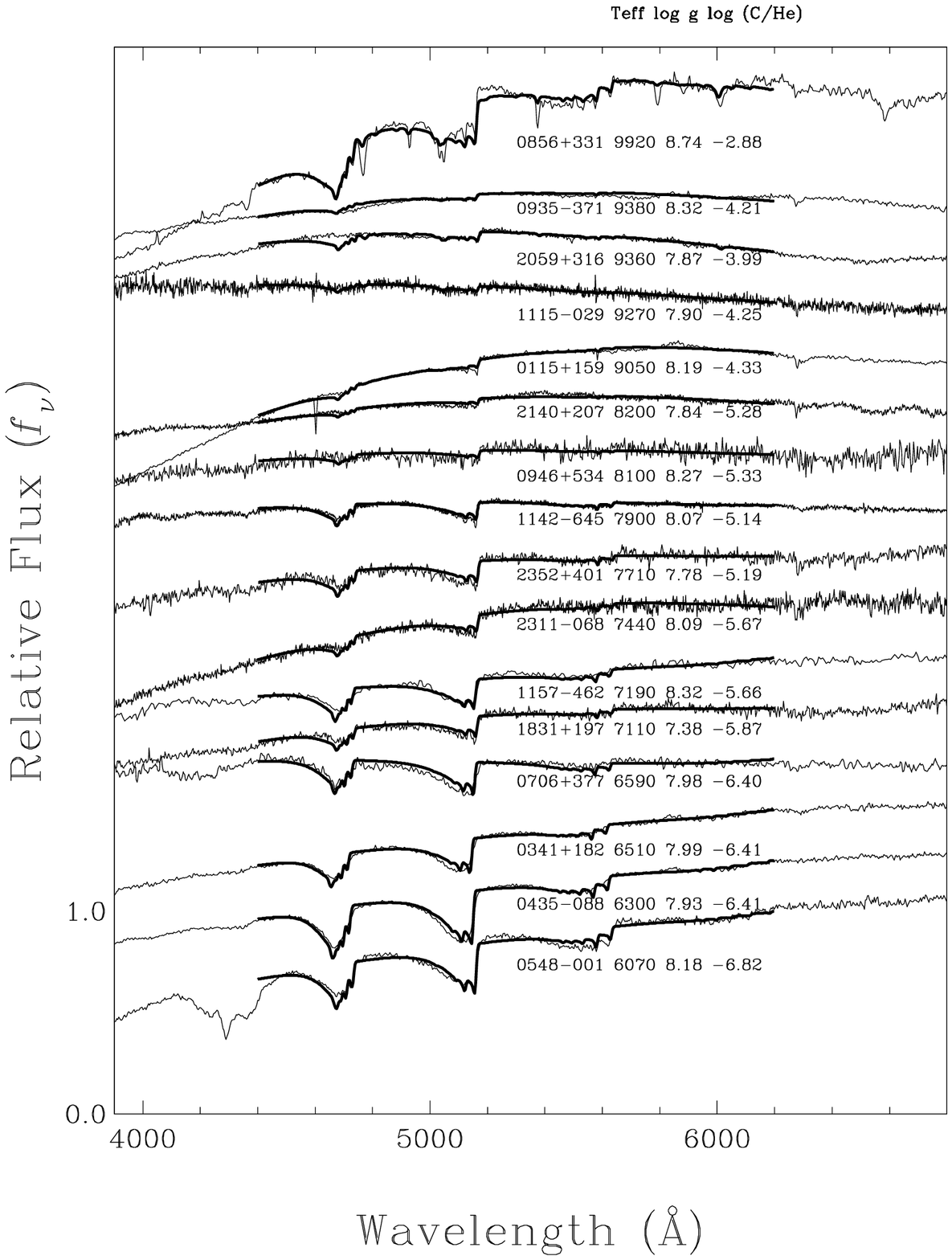] {Fits to the Swan bands of the DQ 
white dwarfs from the BRL/BLR sample, shown in order of
decreasing effective temperatures from top to bottom.  The spectra are
normalized to unity at $6200$ \AA\ and offset from each other by an
arbitrary factor for clarity. WD~0856+331 also shows atomic \ion{C}{1}
lines. \label{fg:f7}}

\figcaption[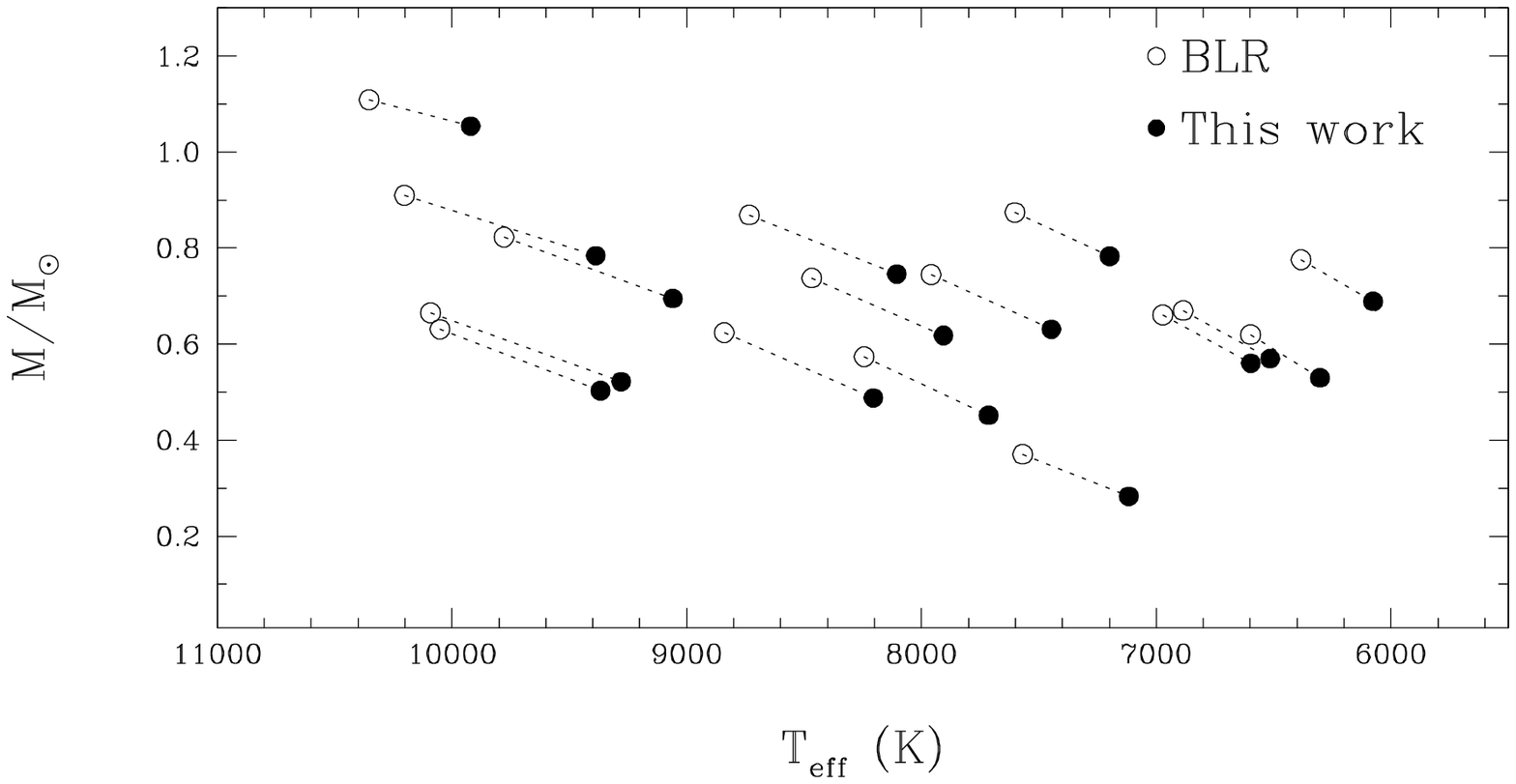] {Comparaison of effective temperatures and stellar masses 
for DQ stars analyzed with pure helium model atmospheres (BLR) and with models
including carbon (this work).\label{fg:f8}}

\figcaption[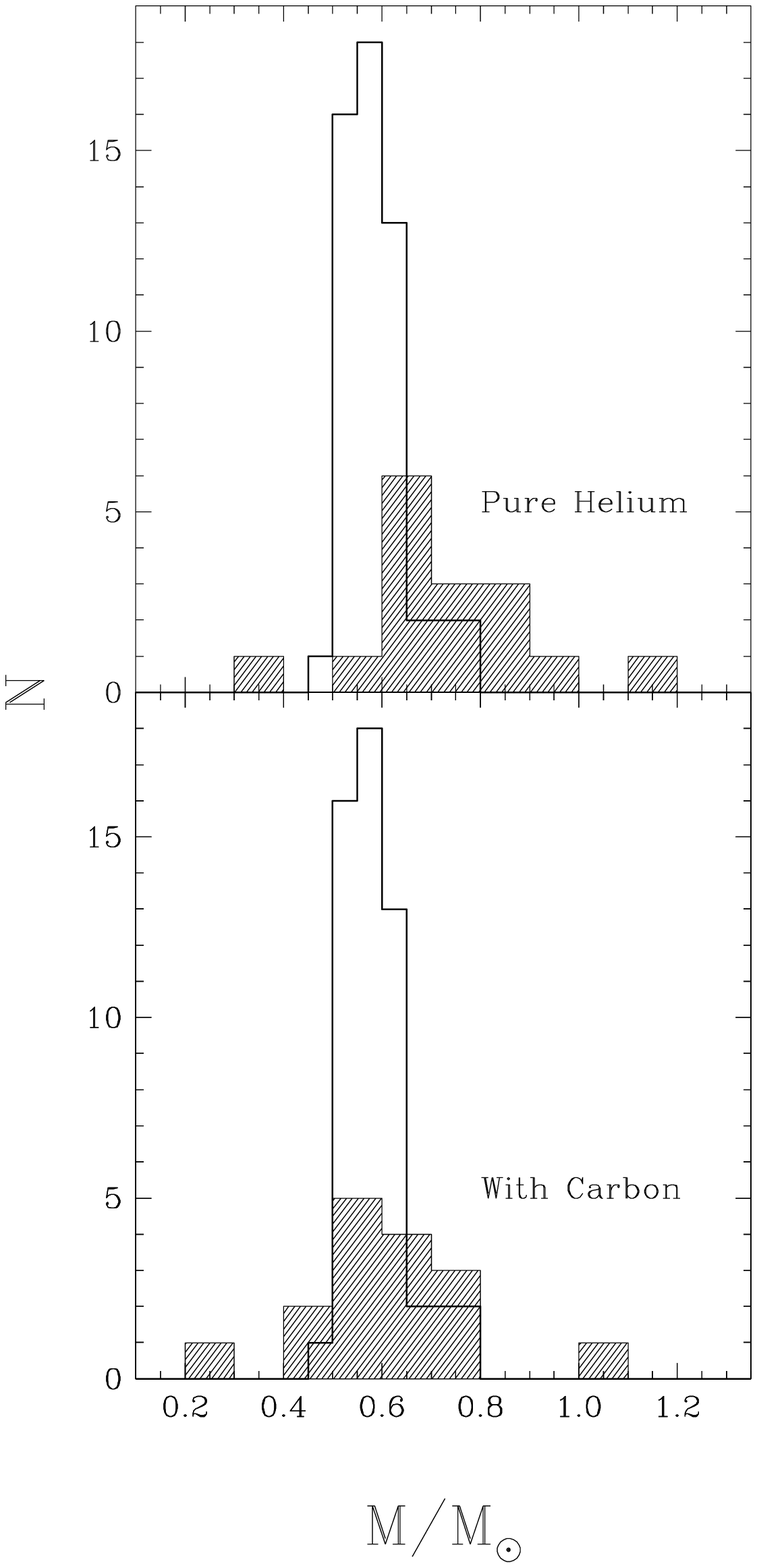] {{\it Top panel:} Mass distribution of DQ stars
analyzed with pure helium atmosphere models ({\it hatched histogram};
$\langle M \rangle = 0.73$ \msun, $\sigma=0.17$~\msun) compared with
the spectroscopic DB mass distribution of \citet{beauchamp96}
({\it solid line}; $\langle M \rangle=0.59$~\msun, $\sigma=0.06$~\msun). 
{\it Bottom panel:} Same as above but with the DQ stars analyzed with models including 
carbon ($\langle M \rangle=0.62$~\msun, $\sigma=0.18$~\msun).\label{fg:f9}}

\figcaption[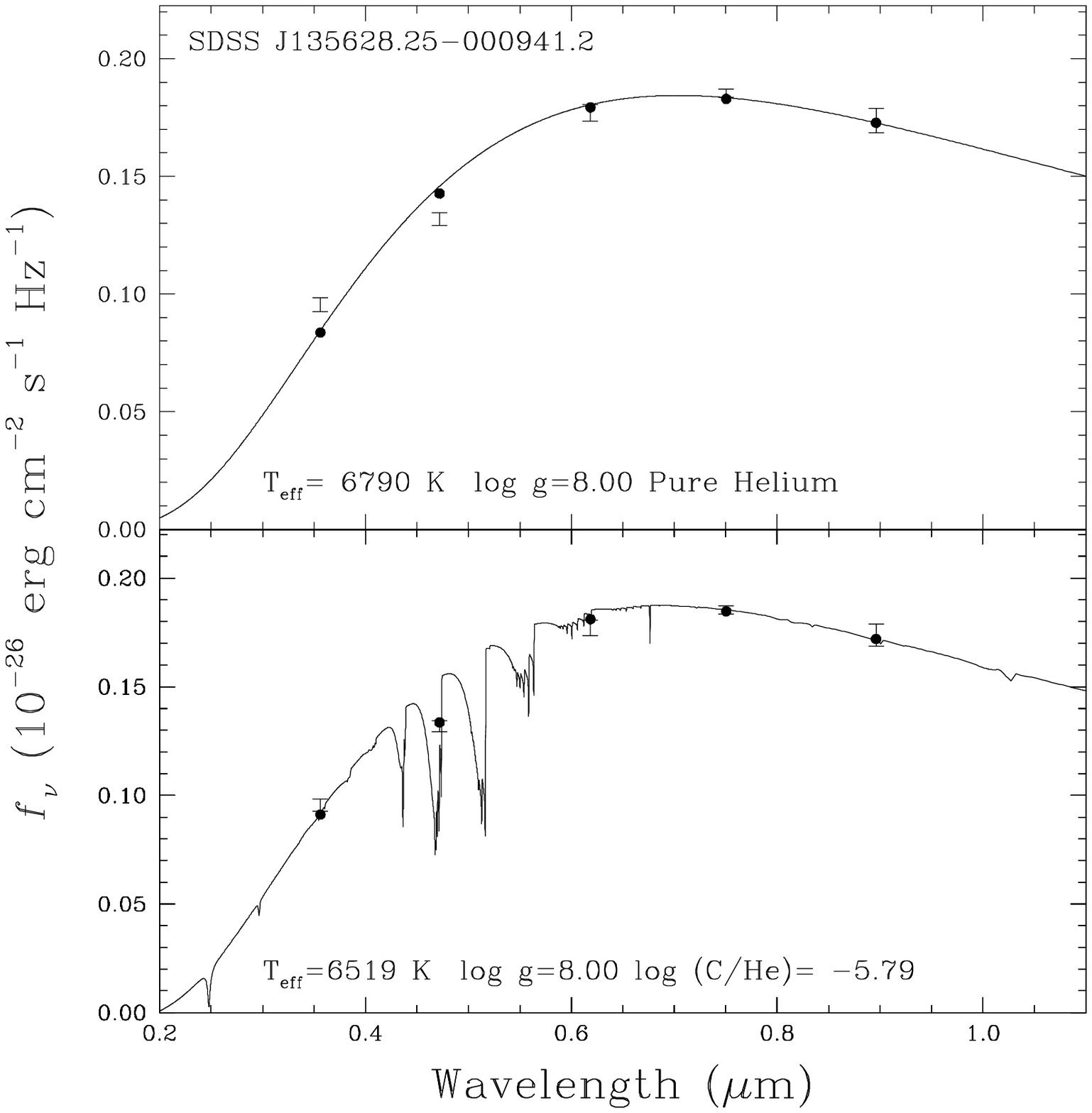] {Fits to the energy distribution
of the DQ star SDSS J135628.25$-$000941.2. The $ugriz$ photometric
observations are represented by errors bars, while the filled circles
correspond to the average model fluxes used in the $\chi^2$ fit. The
top and bottom panels represent the pure helium solution and the
solution including carbon, respectively. The atmospheric parameters
are given in each panel. Also shown are the corresponding
monochromatic Eddington fluxes ({\it solid
lines}).\label{fg:f10}}

\figcaption[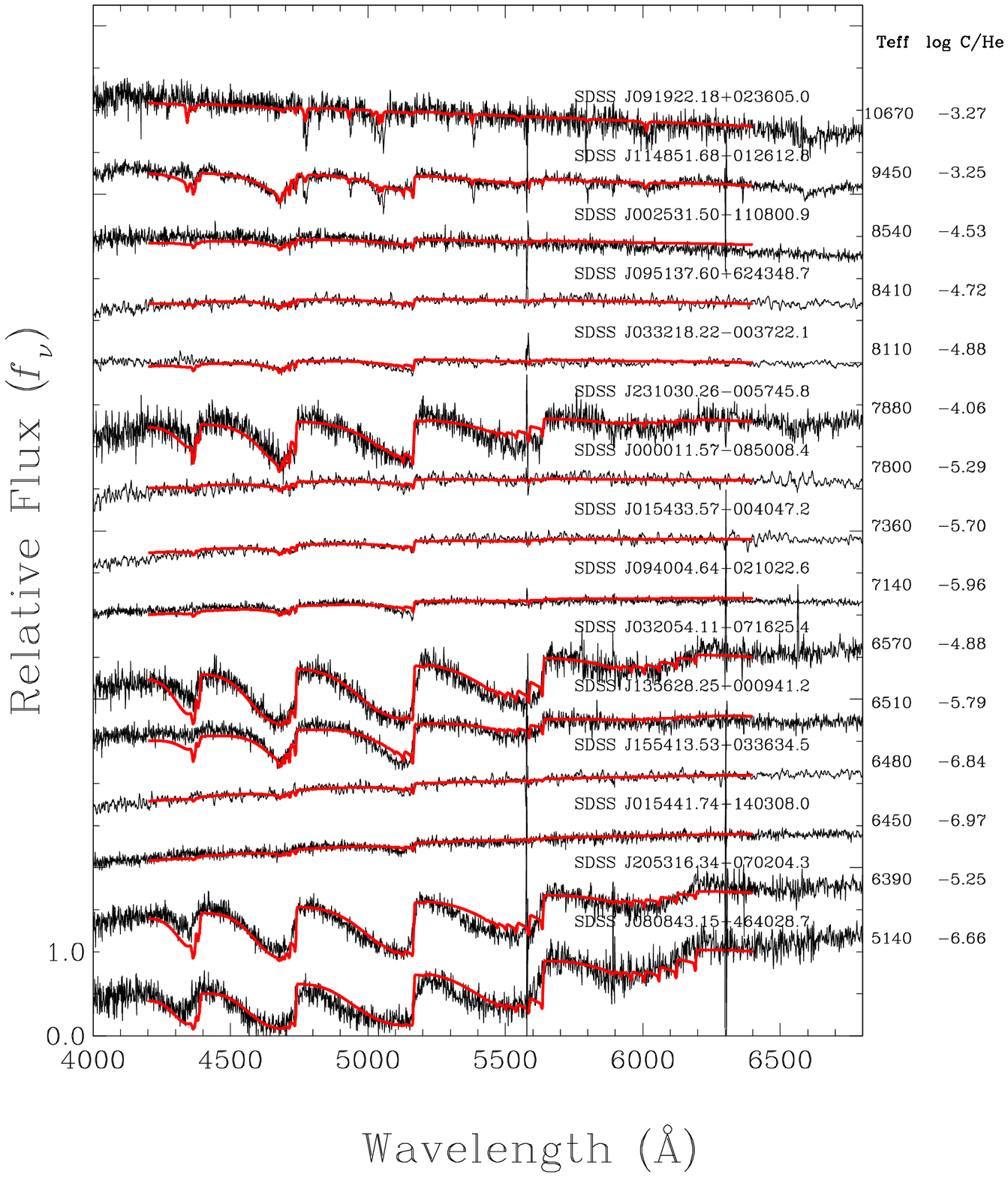] {Sample spectroscopic fits to the Swan bands 
of the SDSS sample plotted in order of decreasing effective
temperatures from top to bottom. The top two stars also show atomic
\ion{C}{1} lines. The spectra are normalized to unity at 6400 \AA\ 
and offset by an arbitrary factor for clarity (see colored version in
the electronic version of this Journal).\label{fg:f11}}

\figcaption[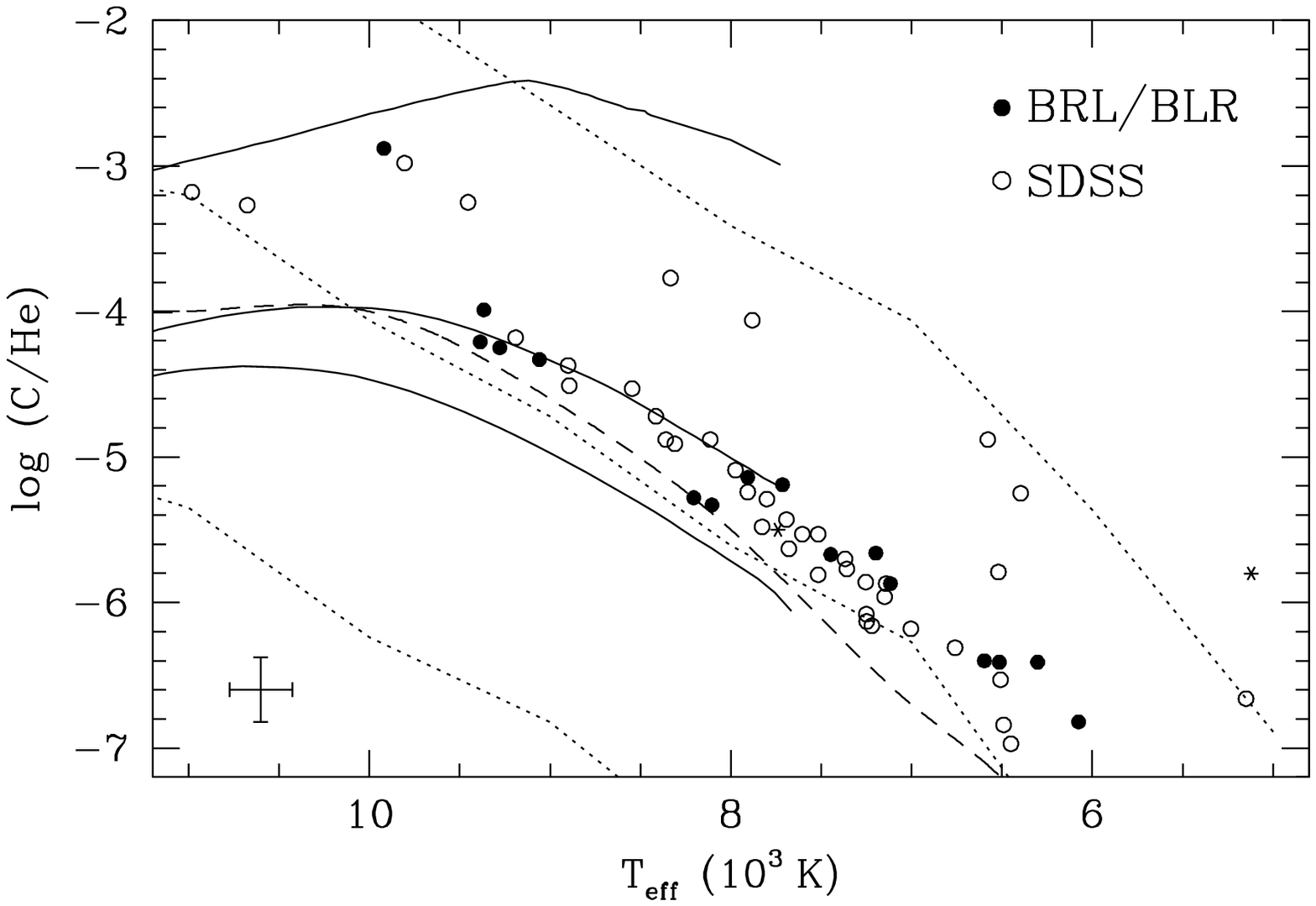] {Measured carbon abundances as a function of
$\Te$ for both the BRL/BLR ({\it filled circles}) and the SDSS ({\it
open circles}) samples; the star symbols show the location of Procyon
B ({\it left symbol}; Provencal et al.~2002) and GSC2U
J131147.2+292348 ({\it right symbol}; Carollo et al.~2003). The error
bars in the lower left corner indicate the average uncertainties of
$\Te$ ($\sim 170$~ K) and $\che$ ($\sim 0.22$ dex). The dashed line
corresponds to the detection threshold of the C$_2$ Swan bands for
$\Te< 10,000$~K, or of the atomic C~\textsc{i} lines for $\Te>
10,000$~K.  The dotted curves represent the evolutionary models of
\citet{pelletier86} at 0.6 \msun\ with, from top to bottom,
$\log q({\rm He})=-4.0$, $-3.5$, and $-3.0$, while the solid curves
represent those of \citet{fon05} at 0.6 \msun\ with, from top to
bottom, $\log q({\rm He})=-4.0$, $-3.0$, and $-2.0$.
\label{fg:f12}}

\figcaption[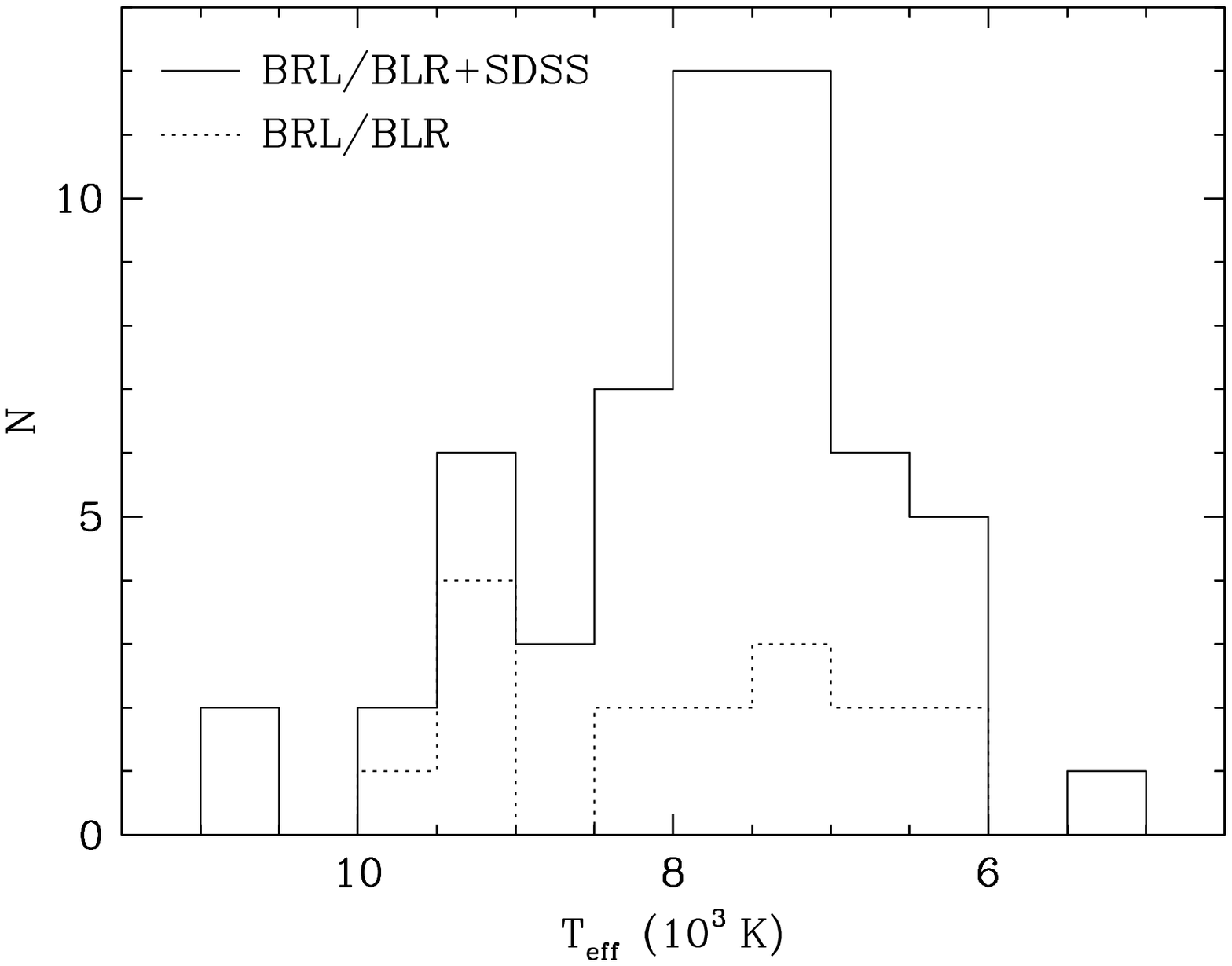] {Number of DQ white dwarfs in the BRL/BLR and SDSS 
samples as a function of effective temperature in 500 K bins.\label{fg:f13}}

\figcaption[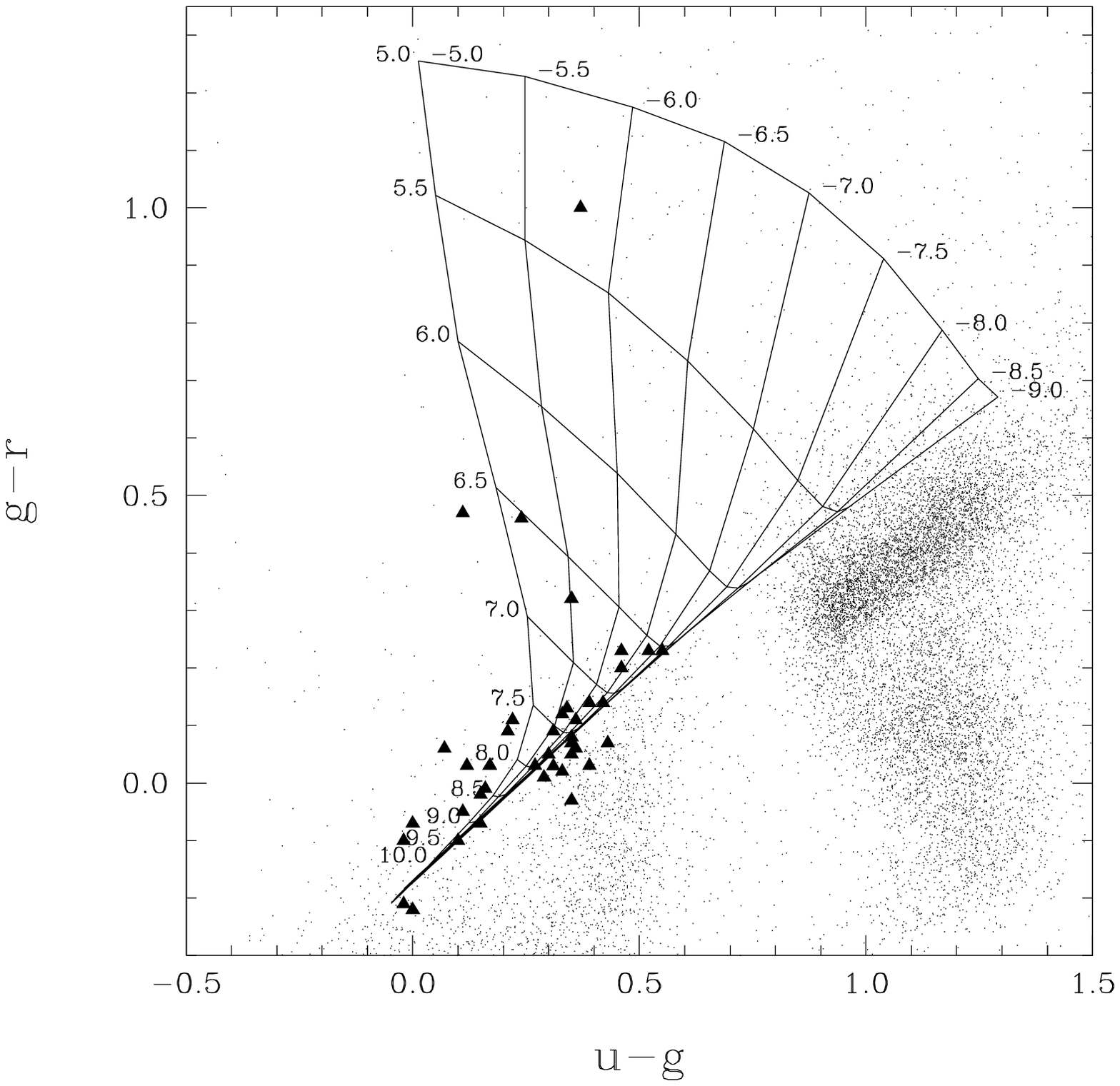] {$u-g$ vs.~$g-r$ color-color diagram. 
Small dots show objects with stellar images while filled triangles 
represent the DQ stars from Table 2. The curves show our DQ
photometric sequences at $\logg=8$ for various values of $\Te$ (in
units of $10^3$ K) and $\che$.\label{fg:f14}}

\clearpage
\begin{figure}[p]
\plotone{f1.eps}
\begin{flushright}
Figure \ref{fg:f1}
\end{flushright}
\end{figure}

\clearpage
\begin{figure}[p]
\plotone{f2.eps}
\begin{flushright}
Figure \ref{fg:f2}
\end{flushright}
\end{figure}

\clearpage
\begin{figure}[p]
\plotone{f3.eps}
\begin{flushright}
Figure \ref{fg:f3}
\end{flushright}
\end{figure}

\clearpage
\begin{figure}[p]
\plotone{f4.eps}
\begin{flushright}
Figure \ref{fg:f4}
\end{flushright}
\end{figure}

\clearpage
\begin{figure}[p]
\plotone{f5.eps}
\begin{flushright}
Figure \ref{fg:f5}
\end{flushright}
\end{figure}

\clearpage
\begin{figure}[p]
\plotone{f6.eps}
\begin{flushright}
Figure \ref{fg:f6}
\end{flushright}
\end{figure}

\clearpage
\begin{figure}[p]
\plotone{f7.eps}
\begin{flushright}
Figure \ref{fg:f7}
\end{flushright}
\end{figure}

\clearpage
\begin{figure}[p]
\plotone{f8.eps}
\begin{flushright}
Figure \ref{fg:f8}
\end{flushright}
\end{figure}

\clearpage
\begin{figure}[p]
\plotone{f9.eps}
\begin{flushright}
Figure \ref{fg:f9}
\end{flushright}
\end{figure}

\clearpage
\begin{figure}[p]
\plotone{f10.eps}
\begin{flushright}
Figure \ref{fg:f10}
\end{flushright}
\end{figure}

\clearpage
\begin{figure}[p]
\plotone{f11.eps}
\begin{flushright}
Figure \ref{fg:f11}
\end{flushright}
\end{figure}

\clearpage
\begin{figure}[p]
\plotone{f12.eps}
\begin{flushright}
Figure \ref{fg:f12}
\end{flushright}
\end{figure}

\clearpage
\begin{figure}[p]
\plotone{f13.eps}
\begin{flushright}
Figure \ref{fg:f13}
\end{flushright}
\end{figure}

\clearpage
\begin{figure}[p]
\plotone{f14.eps}
\begin{flushright}
Figure \ref{fg:f14}
\end{flushright}
\end{figure}

\end{document}